\documentclass[twocolumn,showpacs,superscriptaddress,preprintnumbers,prd]{revtex4}
\usepackage{graphicx,amsmath,dcolumn}
\begin{document}
\preprint{SAGA-HE-197-04}
\title{Nuclear parton distribution functions and their uncertainties}
\author{M. Hirai
        \footnote{Present affiliation: 
                  Institute of Particle and Nuclear Studies, 
                  High Energy Accelerator Research Organization, 
                  1-1, Ooho, Tsukuba, Ibaraki, 305-0801, Japan}}
\email{mhirai@rarfaxp.riken.go.jp}
\affiliation{Radiation Laboratory \\
              RIKEN (The Institute of Physical and Chemical Research) \\
              Wako, Saitama, 351-0198, Japan}
\author{S. Kumano}
\homepage{http://hs.phys.saga-u.ac.jp}
\email{kumanos@cc.saga-u.ac.jp}
\affiliation{Department of Physics, Saga University \\
         Saga, 840-8502, Japan}
\author{T.-H. Nagai}
\email{03sm27@edu.cc.saga-u.ac.jp}
\affiliation{Department of Physics, Saga University \\
         Saga, 840-8502, Japan}
\date{April 8, 2004}
\begin{abstract}
\vspace{0.2cm}
We analyze experimental data of nuclear structure-function ratios
$F_2^A/F_2^{A'}$ and Drell-Yan cross section ratios for obtaining
optimum parton distribution functions (PDFs) in nuclei.
Then, uncertainties of the nuclear PDFs are estimated by the Hessian method.
Valence-quark distributions are determined by the $F_2$ data at large $x$;
however, the small-$x$ part is not obvious from the data.
On the other hand, the antiquark distributions are determined well at
$x \sim 0.01$ from the $F_2$ data and at $x \sim 0.1$ by the Drell-Yan data;
however, the large-$x$ behavior is not clear. Gluon distributions cannot
be fixed by the present data and they have large uncertainties
in the whole $x$ region. Parametrization results are shown in comparison
with the data. We provide a useful code for calculating
nuclear PDFs at given $x$ and $Q^2$.
\end{abstract}
\maketitle

\section{Introduction}\label{intro}
\setcounter{equation}{0}
\vspace{-0.2cm}

Parton distribution functions (PDFs) in the nucleon have been
obtained by analyzing high-energy nucleon reaction data \cite{pdfs}.
Such an analysis is crucial for calculating precise cross sections 
for finding new physics phenomena. These investigations are
valuable for clarifying internal hadron structure, and the studies
ultimately lead to establishment of the nonperturbative aspect of
quantum chromodynamics (QCD).

It is well known that nuclear parton distribution functions (NPDFs)
are modified from those of the nucleon \cite{sumemc}. It was first found
by the European Muon Collaboration (EMC). Now, major features of
the $x$-dependent modification became clear experimentally. Although
a variety of data are not still available in comparison with
the nucleonic case, the PDF parametrization could be done for the NPDFs
\cite{ekrs,saga01,fs03}. The first $\chi^2$ analysis for the NPDFs was
done in Ref. \cite{saga01} by using a similar technique to
the polarized PDF analysis of the Asymmetry Analysis Collaboration (AAC)
\cite{aac00}. There are also related studies on nuclear shadowing
\cite{recent-shadow}. The word ``shadowing" is used for nuclear modification
at $x \lesssim 0.1$ throughout this paper.

These NPDF studies are valuable for describing high-energy nuclear
scattering phenomena \cite{heavy}. High-energy heavy-ion reactions have been
investigated for finding a quark-gluon plasma signature. Because such
a signature should be found in a modification of cross sections,
the NPDFs should be exactly known. In addition, there is a strong demand
from the neutrino community to have precise neutrino-nucleus, typically
the oxygen nucleus, cross sections for investigating neutrino
oscillation phenomena accurately \cite{nuint,sakuda,py}. 
These necessities motivated us to investigate the NPDF parametrization.

In addition, it is interesting to find how well the NPDFs are
determined. There have been studies of PDF uncertainties
in the nucleon. It was investigated in the unpolarized PDFs 
\cite{pdf-error}, and then the studies were extended to the polarized
PDF uncertainties \cite{polpdf-error,aac03}.  Although error bands are
shown for the NPDFs in Ref. \cite{saga01}, they are not based on a rigorous
error analysis. Here, we calculate the NPDF uncertainties by using
the Hessian method, which is a standard statistical procedure
for estimating errors \cite{pdf-error,polpdf-error,aac03}.

The purpose of this paper is to report investigations after
the publication in Ref. \cite{saga01}. In particular, the followings
are added to the previous analysis:
(1) Drell-Yan data are included in the data set.
(2) HERMES data are also added.
(3) Charm-quark distribution is included.
(4) Uncertainties of the NPDFs are estimated by the Hessian method.

This paper consists of the following.
In Sec.\,\ref{method}, the $\chi^2$ analysis method,
in particular the parametrization form and experimental data, is explained.
Analysis results are shown in Sec.\,\ref{results} and they are summarized
in Sec.\,\ref{summary}.

\vspace{-0.1cm}
\section{Analysis method}
\label{method}
\setcounter{equation}{0}
\vspace{-0.2cm}

We discuss the $\chi^2$ analysis method. 
First, the $x$ and $A$ dependence of the initial PDFs is explained,
and comments are given on charm-quark distributions. Then, experimental
data are introduced, and the uncertainty estimation method is explained.

\vspace{0.0cm}
\subsection{Parametrization}
\label{paramet}
\vspace{-0.2cm}

The NPDFs are provided by a number of parameters at a fixed $Q^2$, which
is denoted $Q_0^2$. The NPDFs could be directly expressed by 
a functional form with parameters, which are obtained by
a $\chi^2$ analysis. However, experimental data are not
sufficient for fixing detailed NPDFs. Therefore, it is more practical
at this stage to parametrize nuclear modification rather than
the NPDFs themselves. Namely, a NPDF is taken as
the corresponding nucleonic PDF multiplied by a weight function $w_i$: 
\begin{equation}
f_i^A (x, Q_0^2) = w_i(x,A,Z) \, f_i (x, Q_0^2).
\label{eqn:paramet}
\end{equation}
The nuclear modification part $w_i$ is obtained by a $\chi^2$ analysis.
Here, $A$ is the mass number and $Z$ is the atomic number of a nucleus.

One of the essential points of the $\chi^2$ analysis is how to choose
the $x$ and $A$ dependent functional form. Because nuclear
modification mechanisms are different depending on the $x$ region,
the $A$ dependence could be different in each
$x$ region. If we would like to describe $w_i$ precisely, it could be
a complicated function of mixed $x$ and $A$. 
However, instead of assuming a complicated functional form,
we use a simple one at this stage. We leave such a complicated
analysis for our future work. In Ref. \cite{saga01},
a simple overall $1/A^{1/3}$ dependence is assumed \cite{sd}:
$w_i=1+(1-1/A^{1/3})$($x$ dependent function). Here, we assume the same
functional form. The weight function used for the following analysis
is given by:
\begin{equation}
w_i(x,A,Z)=1+\left( 1 - \frac{1}{A^\alpha} \right) 
          \frac{a_i(A,Z) +b_i x+c_i x^2 +d_i x^3}{(1-x)^{\beta_i}},
\label{eqn:wi}
\end{equation}
where $i$ indicates the parton distribution type, and it is taken
as  $i=u_v$, $d_v$, $\bar q$, and $g$. Among these parameters,
three parameters can be fixed by baryon-number, charge, and momentum
conservations \cite{saga01,fsl}.
The motivation is explained for choosing this functional form 
in Ref. \cite{saga01}.

\subsection{Charm-quark distributions}
\label{charm}
\vspace{-0.2cm}

In the previous analysis, the flavor number is limited to three.
However, charm-quark distributions are important for practical applications.
For example, charmonium productions are used for searching a quark-gluon
plasma signature in heavy-ion reactions. The charm distributions are also
important in neutrino reactions \cite{py}. Therefore, we add nuclear
charm-quark distributions into the analysis.

At $Q^2=m_c^2$, where $m_c$ is the charm-quark mass, 
the running coupling constants for the flavor-number
three and four should agree each other:
$\alpha_s^{N_f=3}(m_c^2) = \alpha_s^{N_f=4}(m_c^2)$.
In the leading order (LO), it leads to the relation between scale parameters:
$\Lambda_3 = \Lambda_4 (m_c/\Lambda_4)^{2/27}$.
Since the initial distributions in Eq. (\ref{eqn:paramet}) are provided
at $Q^2$ which is smaller than $m_c^2$ in our analysis, optimized parameters
for the charm distributions do not exist. The distributions appear simply
as $Q^2$ evolution effects.

\subsection{Experimental data}
\label{data}

\begin{table}[b!]
\vspace{-0.6cm}
\caption{Nuclear species, experiments, references, and the number of
         data points are listed for the used data with $Q^2 \ge 1$ GeV$^2$.}
\label{tab:exp}
\begin{ruledtabular}
\begin{tabular*}{\hsize}
{c@{\extracolsep{0ptplus1fil}}l@{\extracolsep{0ptplus1fil}}c
@{\extracolsep{0ptplus1fil}}c}
nucleus & experiment & reference & \# of data \\
\colrule\colrule
($F_2^A/F_2^D$)
         &           &                  &                    \\
$^4$He/D
         & SLAC-E139 & \cite{slac94}    &    \   18          \\
         & NMC-95    & \cite{nmc95}     &    \   17          \\
Li/D     & NMC-95    & \cite{nmc95}     &    \   17          \\
Be/D     & SLAC-E139 & \cite{slac94}    &    \   17          \\
C/D
         & EMC-88    & \cite{emc88}     &    \ \  9          \\
         & EMC-90    & \cite{emc90}     &    \ \  5          \\
         & SLAC-E139 & \cite{slac94}    &    \ \  7          \\
         & NMC-95    & \cite{nmc95}     &    \   17          \\
         & FNAL-E665-95   & \cite{e665-95} & \ \  5          \\
N/D      
         & BCDMS-85  & \cite{bcdms85}   &    \ \  9          \\
         & HERMES-03 & \cite{hermes03}  &       153          \\
Al/D
         & SLAC-E49  & \cite{slac83B}   &    \   18          \\
         & SLAC-E139 & \cite{slac94}    &    \   17          \\
Ca/D
         & EMC-90    & \cite{emc90}     &    \ \  5          \\
         & NMC-95    & \cite{nmc95}     &    \   16          \\
         & SLAC-E139 & \cite{slac94}    &    \ \  7          \\
         & FNAL-E665-95   & \cite{e665-95} & \ \  5          \\
Fe/D
         & SLAC-E87  & \cite{slac83}    &    \   14          \\
         & SLAC-E140 & \cite{slac88}    &    \   10          \\
         & SLAC-E139 & \cite{slac94}    &    \   23          \\
         & BCDMS-87  & \cite{bcdms87}   &    \   10          \\
Cu/D     & EMC-93    & \cite{emc93}     &    \   19          \\
Kr/D     & HERMES-03 & \cite{hermes03}  &       144          \\
Ag/D     & SLAC-E139 & \cite{slac94}    &    \ \  7          \\
Sn/D     & EMC-88    & \cite{emc88}     &    \ \  8          \\
Xe/D     & FNAL-E665-92   & \cite{e665-92} & \ \  5          \\
Au/D
         & SLAC-E140 & \cite{slac88}    &    \ \  1          \\
         & SLAC-E139 & \cite{slac94}    &    \   18          \\
Pb/D     & FNAL-E665-95   & \cite{e665-95} & \ \  5          \\
\colrule
$F_2^A/F_2^D$ total  &        &         &       606          \\
\colrule
($F_2^A/F_2^{A'}$)
         &           &                  &                    \\
Be/C     & NMC-96    & \cite{nmc96}     &    \   15          \\
Al/C     & NMC-96    & \cite{nmc96}     &    \   15          \\
Ca/C    
         & NMC-95    & \cite{nmc95}     &    \   24          \\
         & NMC-96    & \cite{nmc96}     &    \   15          \\
Fe/C     & NMC-96    & \cite{nmc96}     &    \   15          \\
Sn/C     & NMC-96    & \cite{nmc96snc}  &       146          \\
Pb/C     & NMC-96    & \cite{nmc96}     &    \   15          \\
C/Li     & NMC-95    & \cite{nmc95}     &    \   24          \\
Ca/Li    & NMC-95    & \cite{nmc95}     &    \   24          \\
\colrule
$F_2^{A}/F_2^{A'}$ total  &     &    &       293          \\
\colrule
($\sigma_{DY}^{pA}/\sigma_{DY}^{pA'}$)
         &           &                  &                    \\
C/D      & FNAL-E772-90   & \cite{e772-90}   &    \ \  9     \\
Ca/D     & FNAL-E772-90   & \cite{e772-90}   &    \ \  9     \\
Fe/D     & FNAL-E772-90   & \cite{e772-90}   &    \ \  9     \\
W/D      & FNAL-E772-90   & \cite{e772-90}   &    \ \  9     \\
Fe/Be    & FNAL-E866/NuSea-99 & \cite{e866-99} &  \ \  8     \\
W/Be     & FNAL-E866/NuSea-99 & \cite{e866-99} &  \ \  8     \\
\colrule
Drell-Yan total  &   &                  &    \   52          \\
\colrule\colrule
total            &   &                  &       951  \       \\
\end{tabular*}
\end{ruledtabular}
\end{table}
\normalsize

In the previous version \cite{saga01}, the used experimental data are limited
to the ratios $F_2^A/F_2^D$ where $D$ indicates the deuteron. The data are
from European Muon Collaboration (EMC) \cite{emc88, emc90, emc93}, 
the SLAC-E49, E87, E139, and E140 Collaborations
\cite{slac83,slac83B,slac88,slac94},
the Bologna-CERN-Dubna-Munich-Saclay (BCDMS) Collaboration
\cite{bcdms85, bcdms87}, the New Muon Collaboration (NMC) \cite{nmc95},
and the Fermilab-E665 Collaboration \cite{e665-92,e665-95}.
These data are listed in the $F_2^A/F_2^D$ section of Table \ref{tab:exp}.

In addition to these data, we added HERMES data for the ratios
$F_2^A/F_2^D$, where the nucleus $A$ is for
nitrogen and krypton \cite{hermes03}.
Furthermore, the ratios $F_2^A/F_2^{A'}$ ($A' \ne D$) were measured by
the NMC \cite{nmc95,nmc96,nmc96snc}, and these data are also added. 
The Drell-Yan data taken by the Fermilab-E772 \cite{e772-90}
and E866/NuSea \cite{e866-99} collaborations
are added into the data set for the $\chi^2$ analysis. 
In Refs. \cite{e772-90, e866-99}, $Q^2$ (dimuon mass) values are
not listed. Therefore, we calculated the values in the following way 
\cite{peng}. Relations between the dimuon mass and the target
momentum fraction $x_2$ are listed in Ref. \cite{e772-92}. 
We interpolated these values to obtain the $Q^2$ information.

One may note that HERMES $^3$He data are not included into the data
set. The data are not well reproduced by the present fit, so that 
the data produce a significantly large $\chi^2$ value. It comes from
the fact that the $^3$He is a tightly bound nucleus which cannot
be expressed by the simple $1-1/A^{\alpha}$ dependence. In order to reproduce
such a nucleus, more complicated $A$ dependent function should be used
for the analysis. 

The number of data points is listed in Table \ref{tab:exp}.
The data are for the nuclei: deuteron (D), 
helium-4 ($^4$He), lithium (Li), beryllium (Be), carbon (C),
nitrogen (N), aluminum (Al), calcium (Ca), iron (Fe), copper (Cu),
krypton (Kr), silver (Ag), tin (Sn), xenon (Xe), tungsten (W), gold (Au),
and lead (Pb). The numbers of the $F_2^A/F_2^D$, $F_2^A/F_2^{A'}$ ($A'\ne D$),
and Drell-Yan data are 606, 293, and 52, respectively. The total number
is 951. 

\begin{figure}[b!]
\vspace{-0.3cm}
\includegraphics[width=0.40\textwidth]{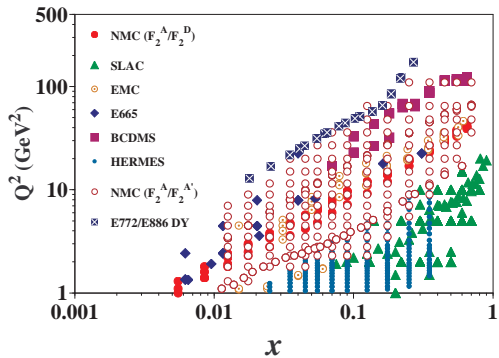}
\vspace{-0.3cm}
\caption{(Color online) Kinematical range is shown by
         $x$ and $Q^2$ values of the used data.}
\label{fig:xq2}
\end{figure}

The kinematical range of the used data is shown in Fig. \ref{fig:xq2}.
The smallest $x$ value with $Q^2 \ge$1 GeV$^2$ is 0.0055 at this stage,
and it is rather limited in comparison with the proton data
($x_{min} \sim 10^{-4}$) at HERA. The SLAC data are taken in the large
$x$, small $Q^2$ region, and the CERN-EMC, NMC, and Fermilab-E665
data are taken in the wide $x$ region from small $x$ to large $x$.
The Drell-Yan data are in the large $Q^2$ region.

\begin{table*}[t!]
\caption{\label{table:parameters}
         Parameters obtained by the analysis.
         The parameters $a_{u_v}$, $a_{d_v}$, and $a_{g}$ are fixed by
         three conservations. Because they depend on nuclear species,
         they are explained separately in Appendix \ref{appen-a}.}
\begin{ruledtabular}
\begin{tabular}{ccccc} 
distribution   & $a$                  & $b$
               & $c$                  & $d$                   \\
\hline
$u_v^A$, $d_v^A$ & fixed (Appendix)     &  2.894 $\pm$  0.395
               & $-$9.390 $\pm$ 1.068   &  7.308 $\pm$  0.866   \\
$\bar q^A$     & $-$0.3794 $\pm$ 0.0461 &  8.626 $\pm$  1.551
               & $-$56.64  $\pm$ 11.84  & 94.11  $\pm$ 27.57    \\
$g^A$          & fixed (Appendix)       &  2.165 $\pm$  3.126 
               & 0.000 (fixed)          &  1.349 $\pm$ 44.56
\end{tabular}
\end{ruledtabular}
\end{table*}

\subsection{$\chi^2$ analysis}
\label{chi2}

Nuclear modification of the PDFs is expressed by the weight functions
$w_i$. We introduce four types by assuming the flavor symmetric
antiquark distributions ($\bar u^A=\bar d^A=\bar s^A\equiv \bar q^A$)
at $Q_0^2$:
\begin{align}
u_v^A (x,Q_0^2) & = w_{u_v} (x,A,Z) \, \frac{Z u_v (x,Q_0^2) 
                                         + N d_v (x,Q_0^2)}{A},
\nonumber \\
d_v^A (x,Q_0^2) & = w_{d_v} (x,A,Z) \, \frac{Z d_v (x,Q_0^2) 
                                         + N u_v (x,Q_0^2)}{A},
\nonumber \\
\bar q^A (x,Q_0^2) & = w_{\bar q} (x,A,Z) \, \bar q (x,Q_0^2),
\nonumber \\
g^A (x,Q_0^2)      & = w_{g} (x,A,Z) \, g (x,Q_0^2).
\label{eqn:wpart}
\end{align}
In the first two equations, the $Z$ terms indicate the proton contributions
and the $N$ terms indicate the neutron ones if there were no nuclear
modification and isospin symmetry could be applied.
Although the antiquark distributions ($\bar u$, $\bar d$, $\bar s$)
in the nucleon are different \cite{flavor3}, there is no clear data
which indicates the difference in nuclei at this stage. Therefore,
the flavor symmetric antiquark distributions are assumed.
The initial scale is chosen $Q_0^2$=1 GeV$^2$. 
The MRST01-LO (Martin-Roberts-Stirling-Thorne, leading-order version of
2001) parametrization \cite{mrst01} is used for the PDFs, so that scale
parameter $\Lambda$ and charm-quark mass $m_c$ are the MRST01 values
in the following analysis.

Using these NPDFs, we calculate the structure-function ratios
$F_2^A/F_2^{A'}$ and the Drell-Yan cross section ratios 
$\sigma_{DY}^{pA}/\sigma_{DY}^{pA'}$ in the leading order (LO)
of $\alpha_s$. The NPDFs are given at $Q_0^2$ in
Eq. (\ref{eqn:wpart}), so that they are evolved to the experimental
$Q^2$ points by the Dokshitzer-Gribov-Lipatov-Altarelli-Parisi (DGLAP)
evolution equations in order to calculate these ratios.
The total $\chi^2$ is defined by 
\begin{equation}
\chi^2 = \sum_j \frac{(R_{j}^{data}-R_{j}^{theo})^2}
                     {(\sigma_j^{data})^2},
\label{eqn:chi2}
\end{equation}
where $R_j$ indicates that the ratios, $F_2^A/F_2^{A'}$ and 
$\sigma_{DY}^{pA}/\sigma_{DY}^{pA'}$.
The experimental errors are calculated from
systematic and statistical errors by
$(\sigma_j^{data})^2 = (\sigma_j^{sys})^2 + (\sigma_j^{stat})^2$.
The optimization of the NPDFs is done by the CERN subroutine
{\tt MINUIT} \cite{minuit}.

\subsection{Uncertainty of nuclear PDFs}
\label{error}

Because the situation of the NPDFs is not as good as the one of
the PDFs in the nucleon, it is especially important to show the
reliability of obtained NPDFs. The uncertainties are shown
in the previous version \cite{saga01}; however, they are 
simply estimated by shifting each parameter by the amount
of the error. Of course, a standard error analysis is needed
for the NPDFs by taking into account correlations among
the parameter errors.

One of the popular ways is to use the Hessian method. In fact, it is
used for the unpolarized PDF analysis of the nucleon \cite{pdf-error}
and also for the polarized PDFs \cite{polpdf-error,aac03}.
Because the method is discussed in Ref. \cite{aac03}, we
explain only a brief outline. 

The parameters of the initial NPDFs in Eq. (\ref{eqn:wi})
are denoted $\xi_i$ ($i$=1, 2, $\cdot \cdot \cdot$, $N$), where $N$
is the number of the parameters.
The $\chi^2$ could be expanded around the minimum point $\hat \xi$:
\begin{equation}
        \Delta \chi^2 \equiv \chi^2(\hat{\xi}+\delta \xi)
          -\chi^2(\hat{\xi})
        =\sum_{i,j} H_{ij}\delta \xi_i \delta \xi_j \, ,
        \label{eq:dchi2}
\end{equation}
where $H_{ij}$ is called Hessian.
The details are discussed elsewhere for the uncertainty estimation 
of the PDFs by the Hessian method. For the detailed explanation, 
one may read Refs. \cite{pdf-error,aac03} about $\Delta \chi^2$
and the Hessian.
A confidence region is identified by providing the $\Delta \chi^2 $ value,
which is determined in the following way.
The confidence level $P$ could be chosen as the one-$\sigma$-error range
of the normal distribution ($P=0.6826$). For one parameter,
$P=0.6826$ is obtained with $\Delta \chi^2$=1. However, 
a different value should be assigned for the $N$ degrees of freedom
\cite{aac03}. For example, if there are nine parameters,
the $\Delta \chi^2$ value is calculated as $\Delta \chi^2=10.427$.

The uncertainty of a NPDF $f^A(x,\hat{\xi})$ is calculated by
the Hessian matrix, which is obtained by running the {\tt MINUIT}
subroutine, and derivatives of the distribution:
\begin{equation}
        [\delta f^A(x)]^2=\Delta \chi^2 \sum_{i,j}
          \left( \frac{\partial f^A(x,\hat{\xi})}{\partial \xi_i}  \right)
          H_{ij}^{-1}
          \left( \frac{\partial f^A(x,\hat{\xi})}{\partial \xi_j}  \right) 
\, .
        \label{eq:dnpdf}
\end{equation}
The derivatives are calculated analytically at the initial scale $Q^2_0$,
and then they are evolved to certain $Q^2$ by the DGLAP evolution equations.

\section{Results}
\label{results}
\setcounter{equation}{0}

Analysis results are discussed. 
First, optimized parameters are shown, and $\chi^2$ contributions from
nuclear data sets are listed. Then, fit results are compared with
experimental data. The actual NPDFs and their uncertainties 
are shown for some nuclei at $Q_0^2$.

\subsection{Comparison with $x$-dependent data}
\label{comp}

\begin{table}[b!]
\caption{Each $\chi^2$ contribution.}
\label{tab:chi2}
\begin{ruledtabular}
\begin{tabular*}{\hsize}
{c@{\extracolsep{0ptplus1fil}}c@{\extracolsep{0ptplus1fil}}c}
nucleus  & \# of data & $\chi^2$  \\
\colrule
$^4$He/D &   \   35    &  \     56.0         \\
Li/D     &   \   17    &  \     88.7         \\ 
Be/D     &   \   17    &  \     44.1         \\
C/D      &   \   43    &       130.8         \\
N/D      &   \  162    &       136.9         \\
Al/D     &   \   35    &  \     43.1         \\
Ca/D     &   \   33    &  \     42.0         \\
Fe/D     &   \   57    &  \     95.7         \\
Cu/D     &   \   19    &  \     11.8         \\
Kr/D     &      144    &       126.9         \\
Ag/D     &   \ \  7    &  \     12.8         \\
Sn/D     &   \ \  8    &  \     14.6         \\
Xe/D     &   \ \  5    &  \ \    2.0         \\
Au/D     &   \   19    &  \     61.6         \\
Pb/D     &   \ \  5    &  \ \    5.6         \\	
\colrule
$F_2^A/F_2^D$ total
         &      606    &       872.8         \\
\colrule
Be/C     &   \   15    &  \     16.1         \\
Al/C     &   \   15    &  \ \    6.1         \\
Ca/C     &   \   39    &  \     36.5         \\
Fe/C     &   \   15    &  \     10.3         \\
Sn/C     &      146    &       257.3         \\
Pb/C     &   \   15    &  \     25.3         \\
C/Li     &   \   24    &  \     78.1         \\
Ca/Li    &   \   24    &       107.7         \\
\colrule
$F_2^{A_1}/F_2^{A_2}$ total 
         &      293    &       537.4         \\
\colrule
C/D      &   \ \  9    &  \ \    9.8         \\
Ca/D     &   \ \  9    &  \ \    7.2         \\
Fe/D     &   \ \  9    &  \ \    8.1         \\
W/D      &   \ \  9    &  \     18.3         \\
Fe/Be    &   \ \  8    &  \ \    6.5         \\
W/Be     &   \ \  8    &  \     29.6         \\
\colrule
Drell-Yan total  
         &   \   52    &  \     79.6         \\
\colrule\colrule
total    &      951    &       1489.8 \      \\
\end{tabular*}
\end{ruledtabular}
\end{table}
\vspace{+0.0cm}
\normalsize

\begin{figure}[h]
        \includegraphics*[width=40mm]{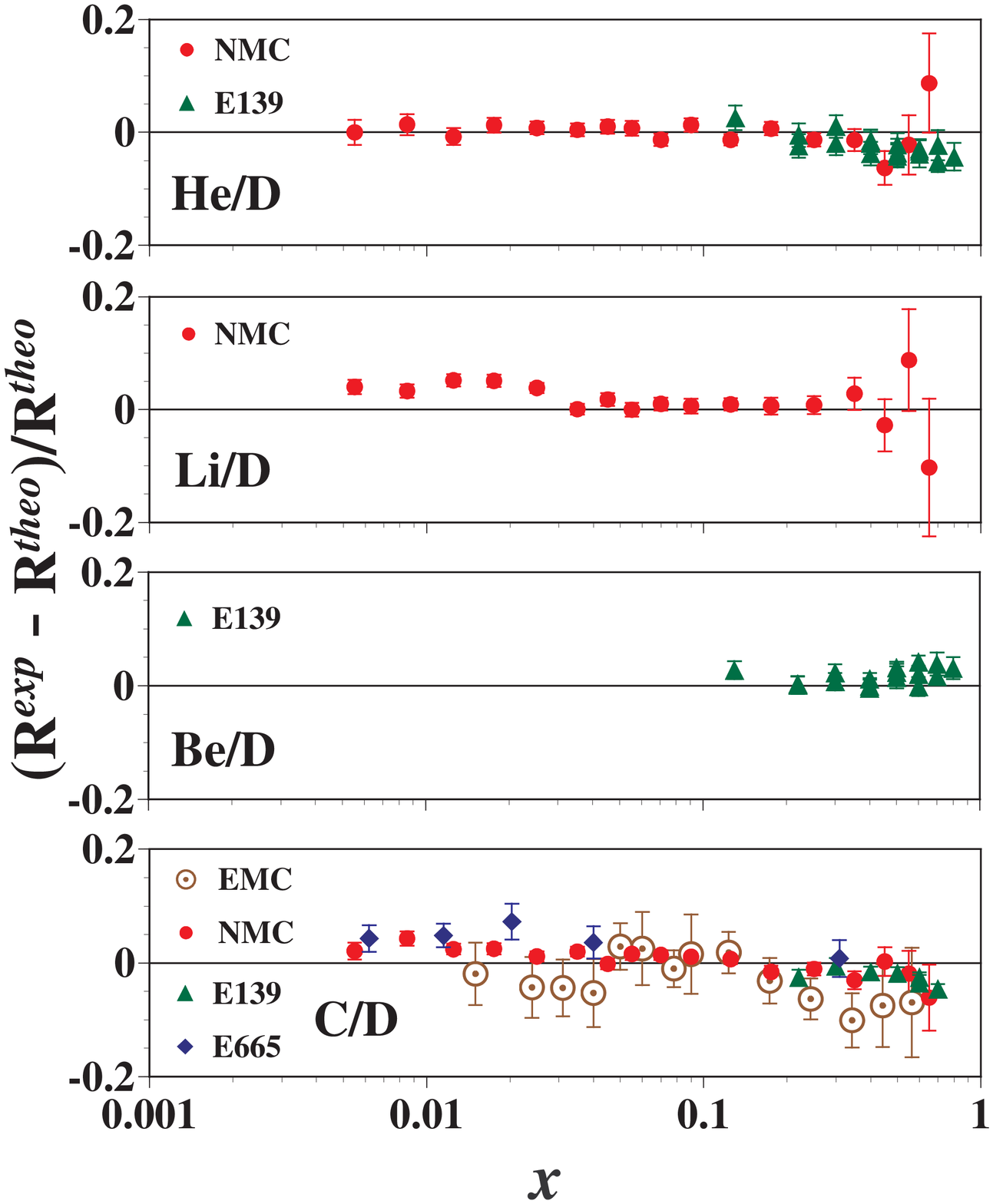} \hspace{0.5mm}
        \includegraphics*[width=37mm]{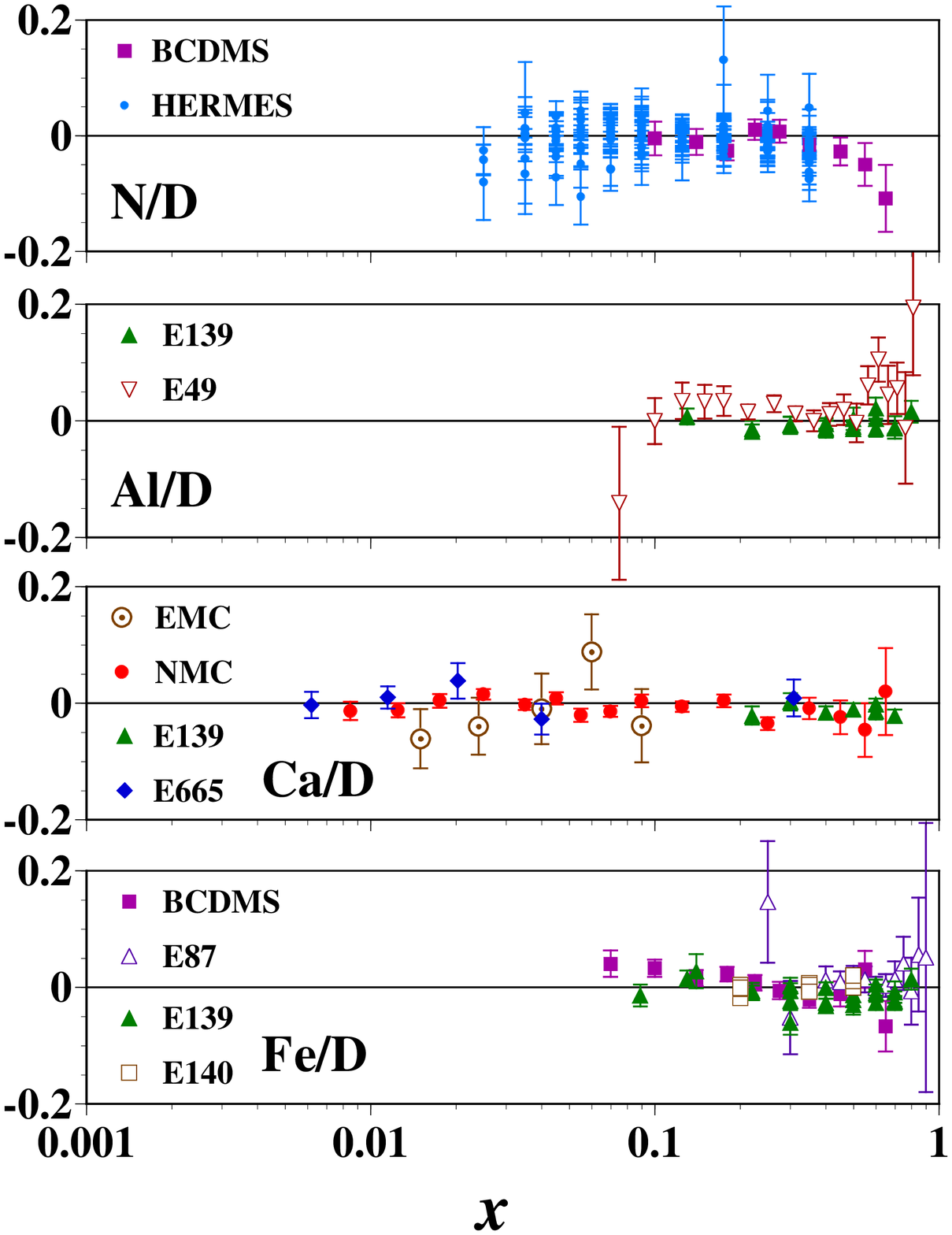} \\
\vspace{0.5cm}
        \includegraphics*[width=40mm]{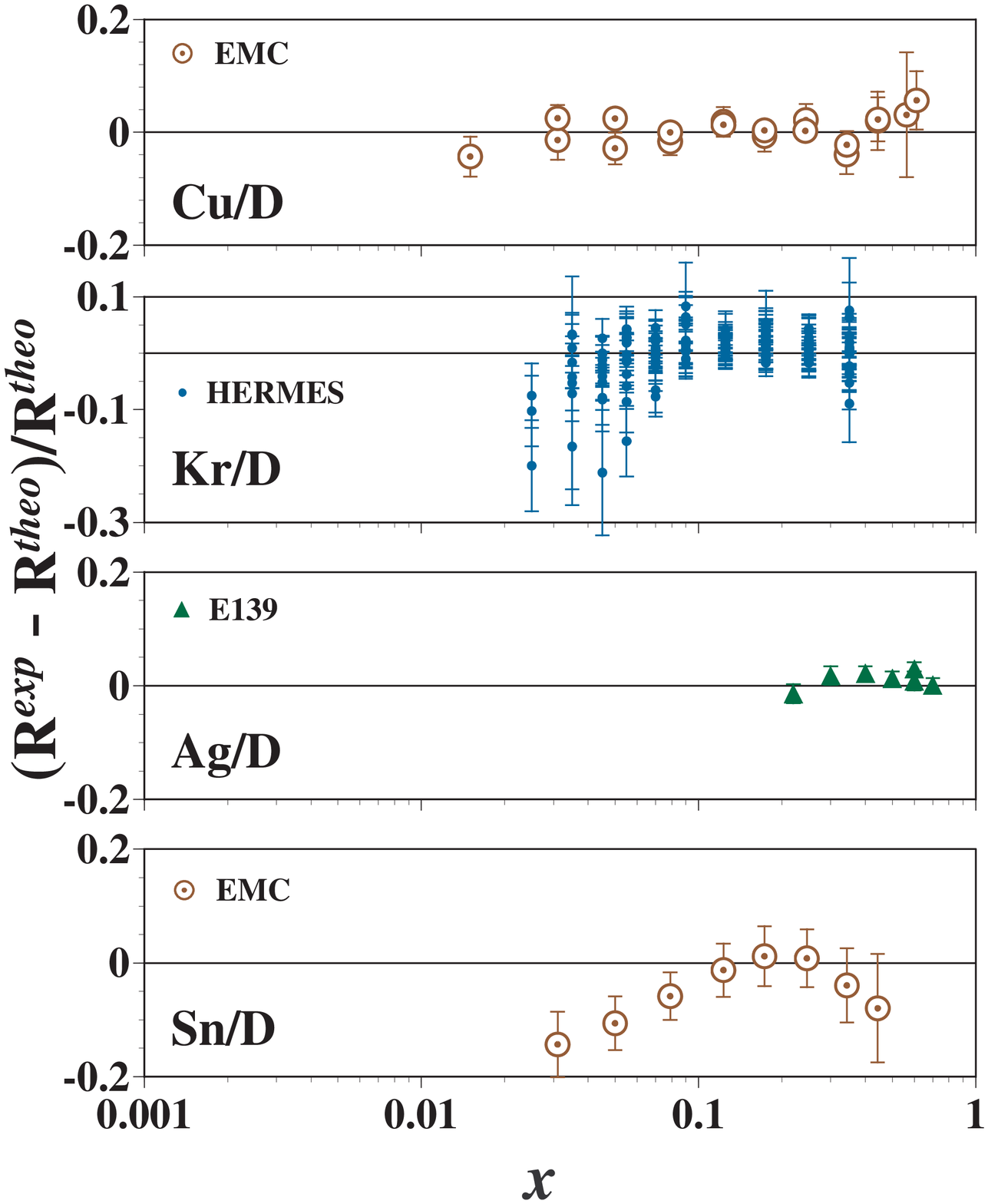} \hspace{0.5mm}
        \includegraphics*[width=37mm]{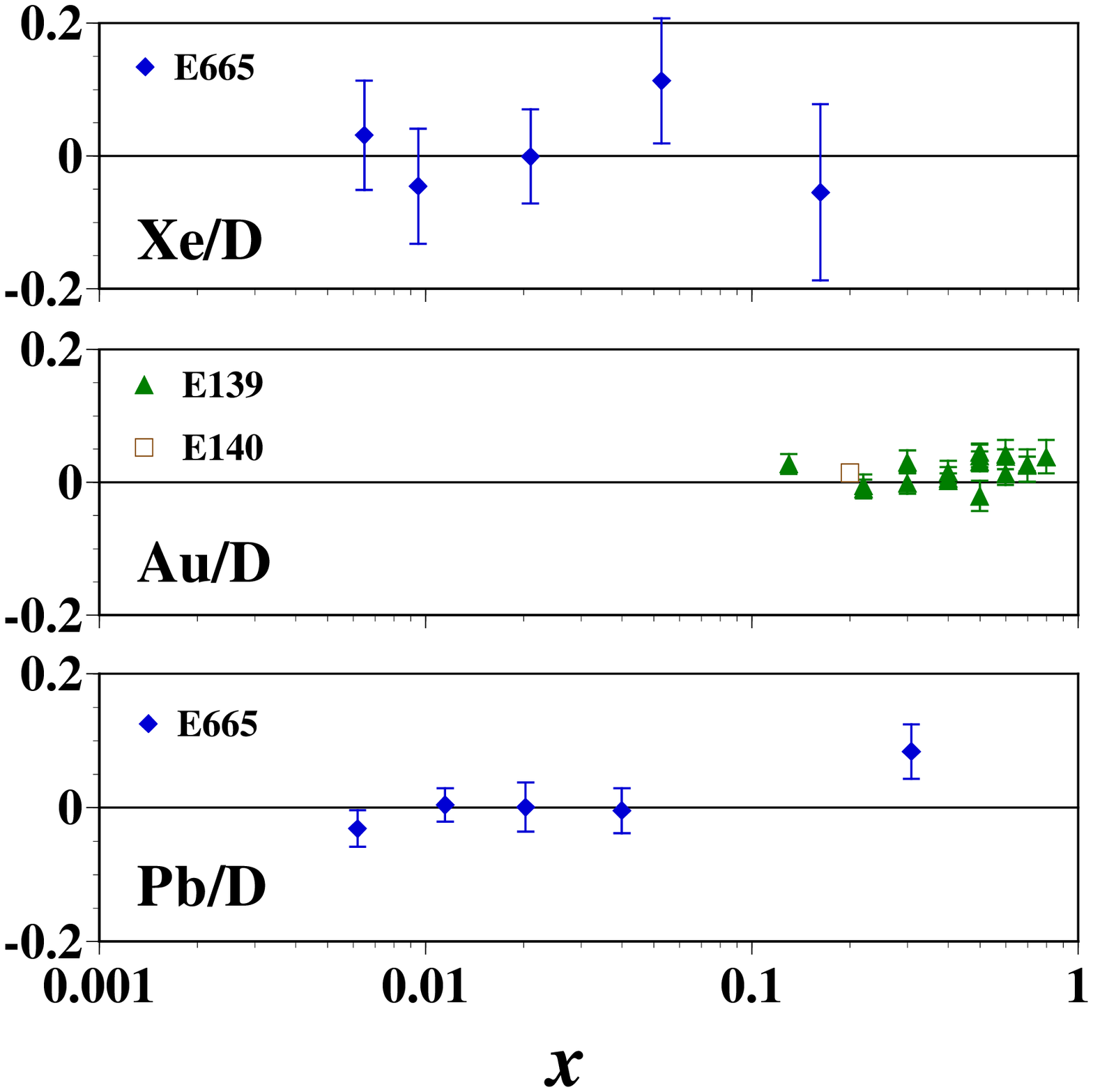} 
       \vspace{-0.25cm}
\caption{\label{fig:RD}
(Color online) Comparison with experimental ratios $R=F_2^A/F_2^D$. 
The ordinate indicates the fractional differences between
experimental data and theoretical values: $(R^{exp}-R^{theo})/R^{theo}$.
}
\end{figure}
\vspace{-0.0cm}
\begin{figure}[h]
        \includegraphics*[width=40mm]{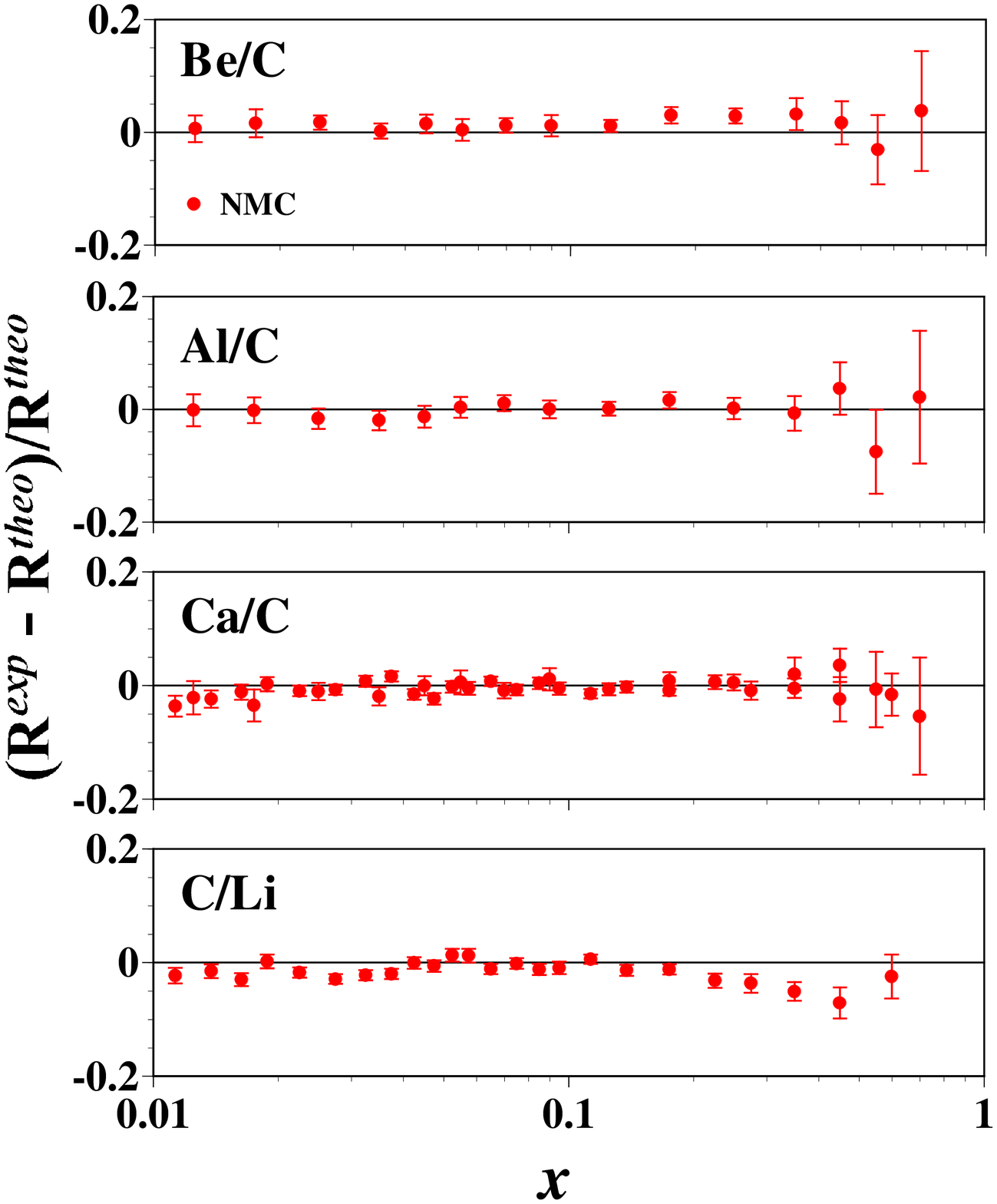} \hspace{0.5mm}
        \includegraphics*[width=37mm]{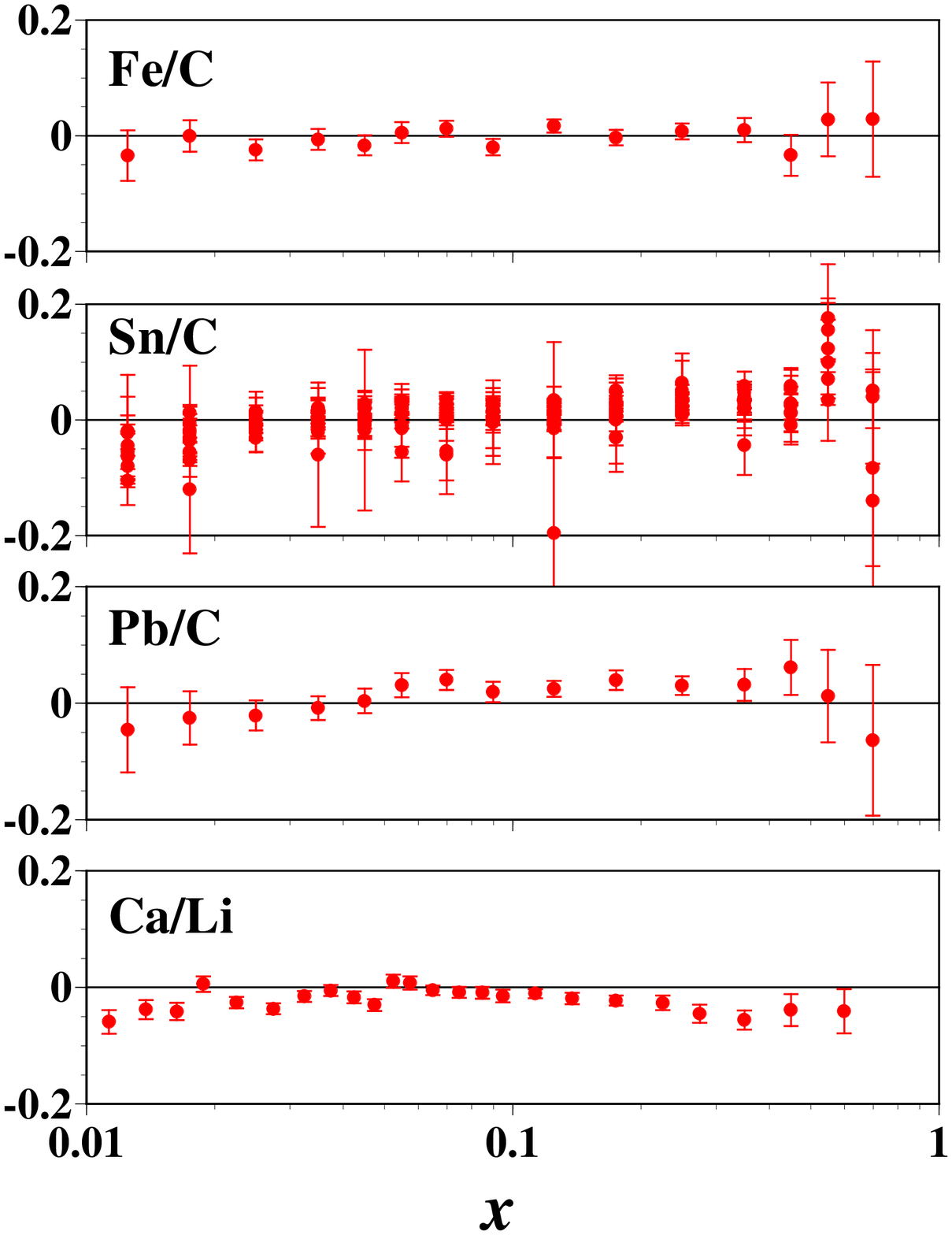} \\
       \vspace{-0.25cm}
\caption{\label{fig:RA}
(Color online) Comparison with experimental data of $R=F_2^A/F_2^{C,Li}$. 
The ratios $(R^{exp}-R^{theo})/R^{theo}$ are shown.
}
\vspace{-0.3cm}
\end{figure}
\vspace{-0.0cm}
\begin{figure}[t]
        \includegraphics*[width=40mm]{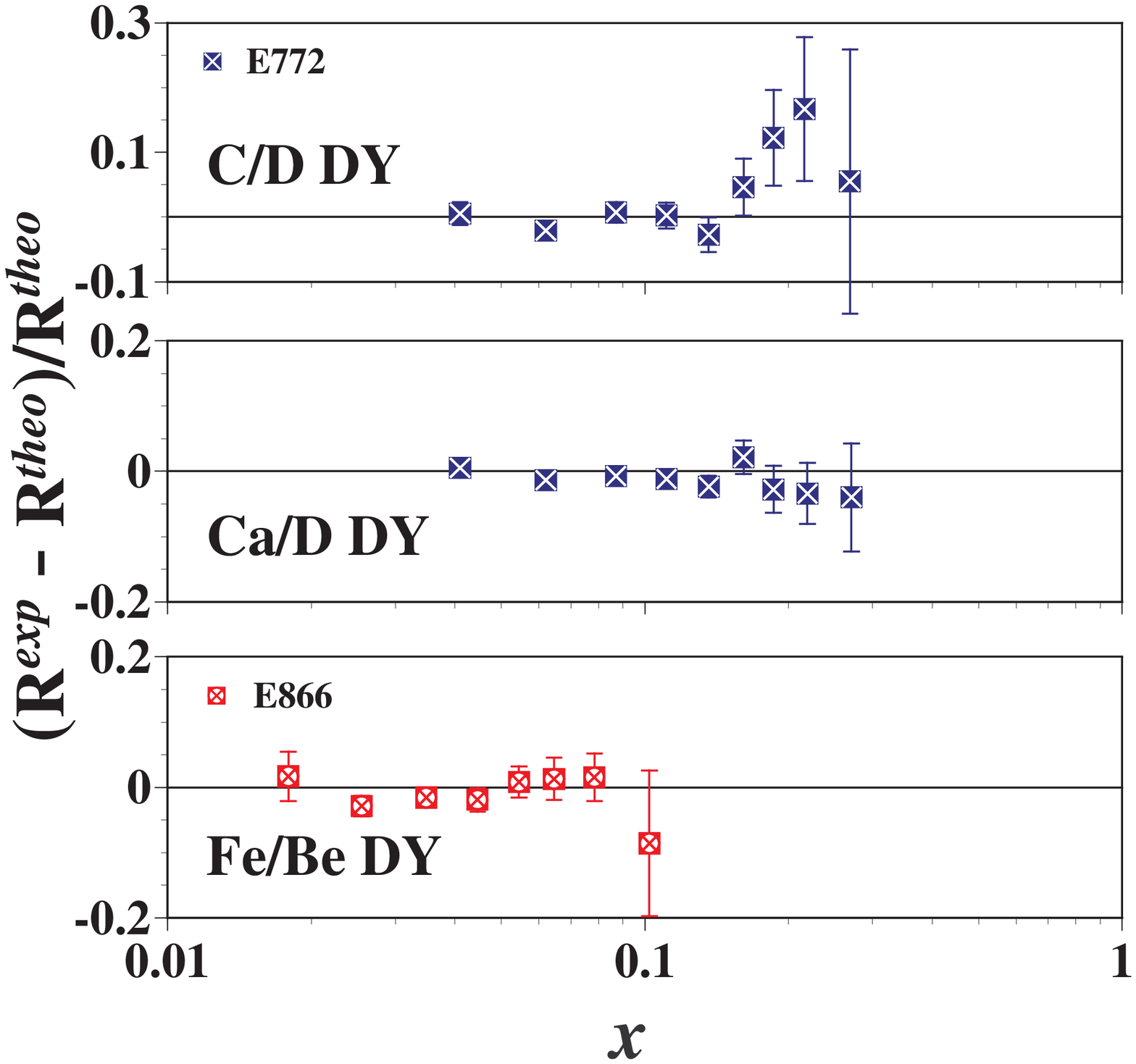} \hspace{0.5mm}
        \includegraphics*[width=37mm]{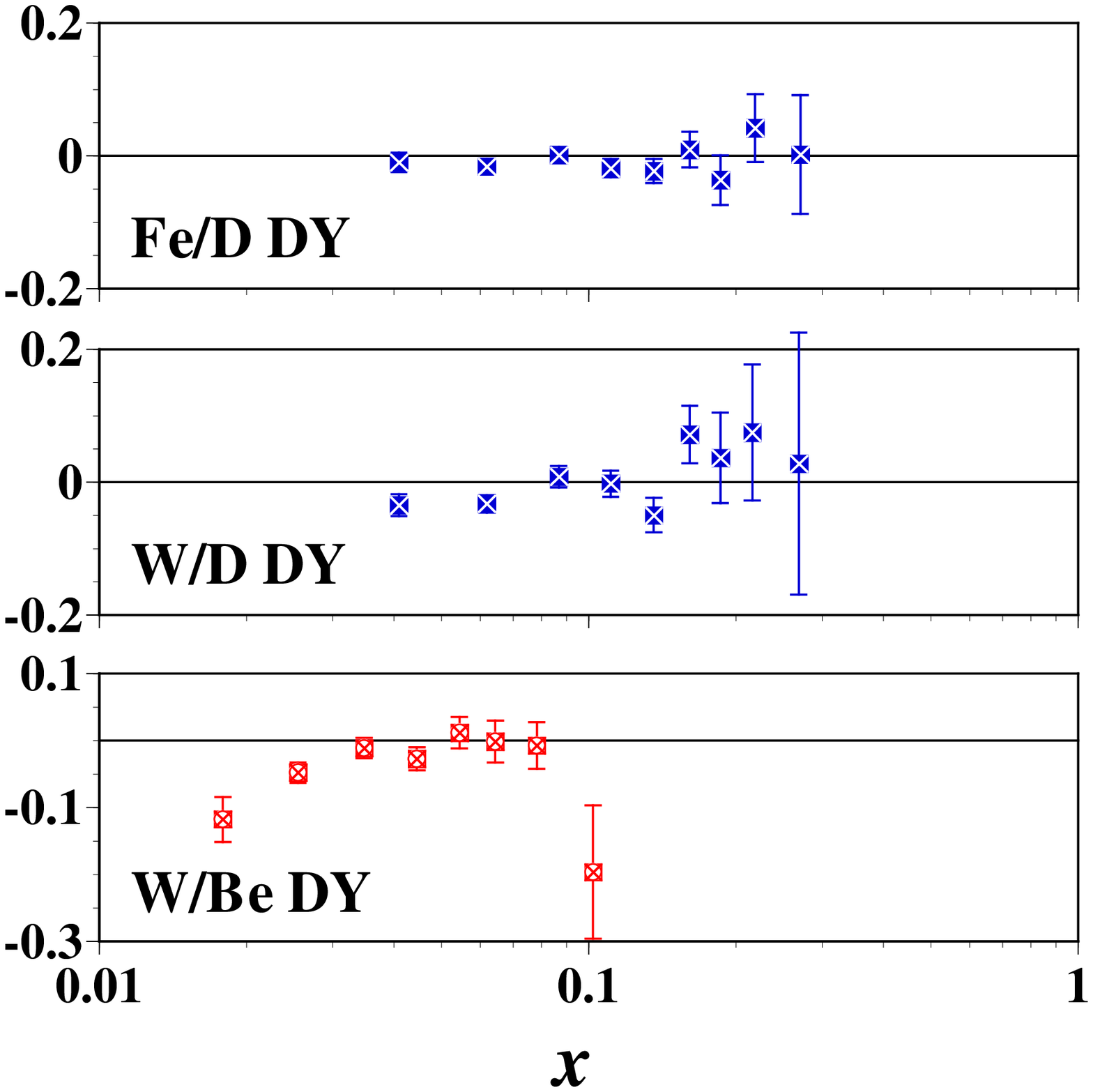} \\
       \vspace{-0.25cm}
\caption{\label{fig:DY}
(Color online) Comparison with Drell-Yan data of
$R=\sigma_{DY}^{pA}/\sigma_{DY}^{pA'}$.
The ratios $(R^{exp}-R^{theo})/R^{theo}$ are shown.
}
\end{figure}
\vspace{-0.0cm}
\begin{figure}[h]
        \includegraphics*[width=60mm]{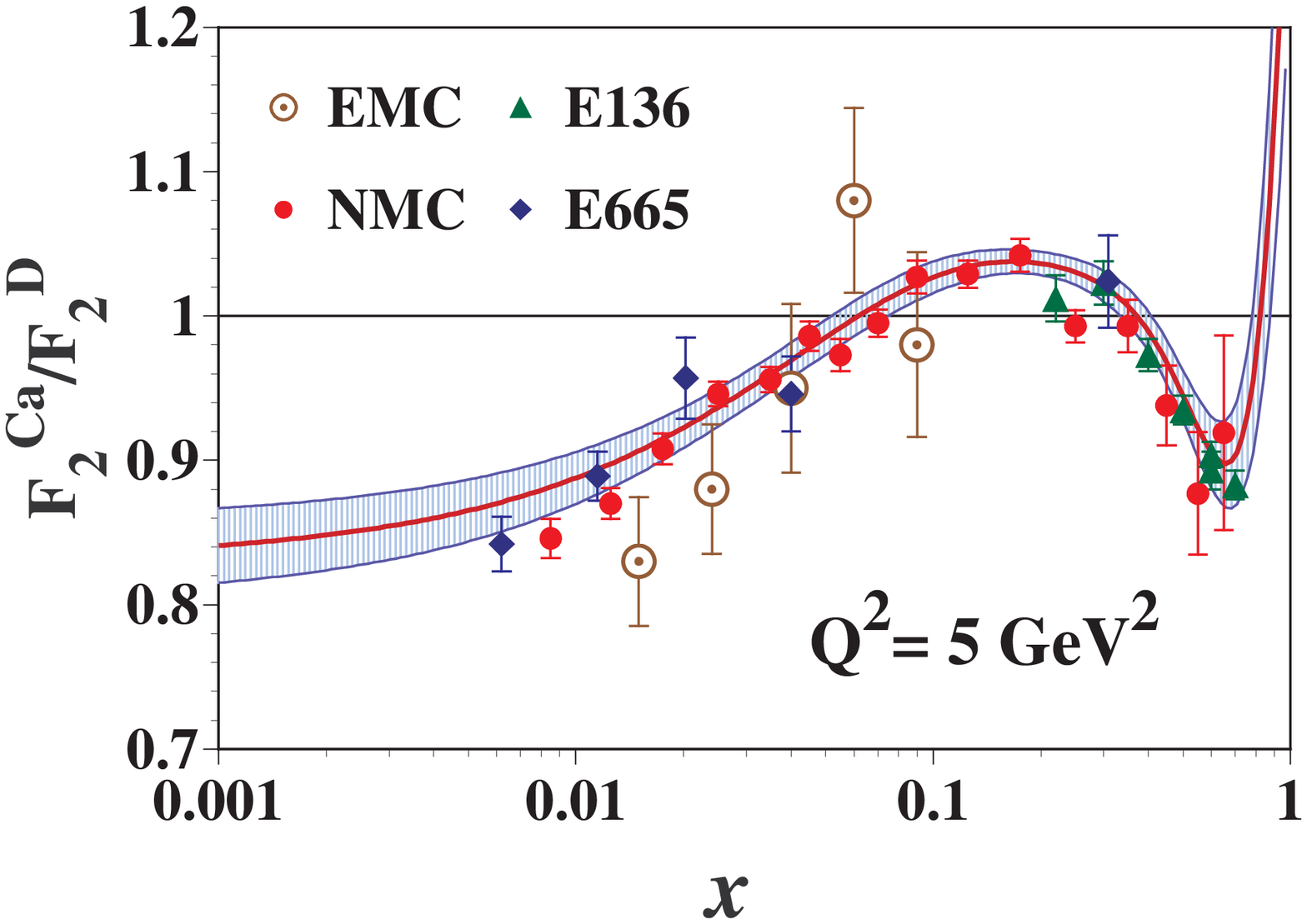} \\
        \vspace{+0.1cm}
        \includegraphics*[width=60mm]{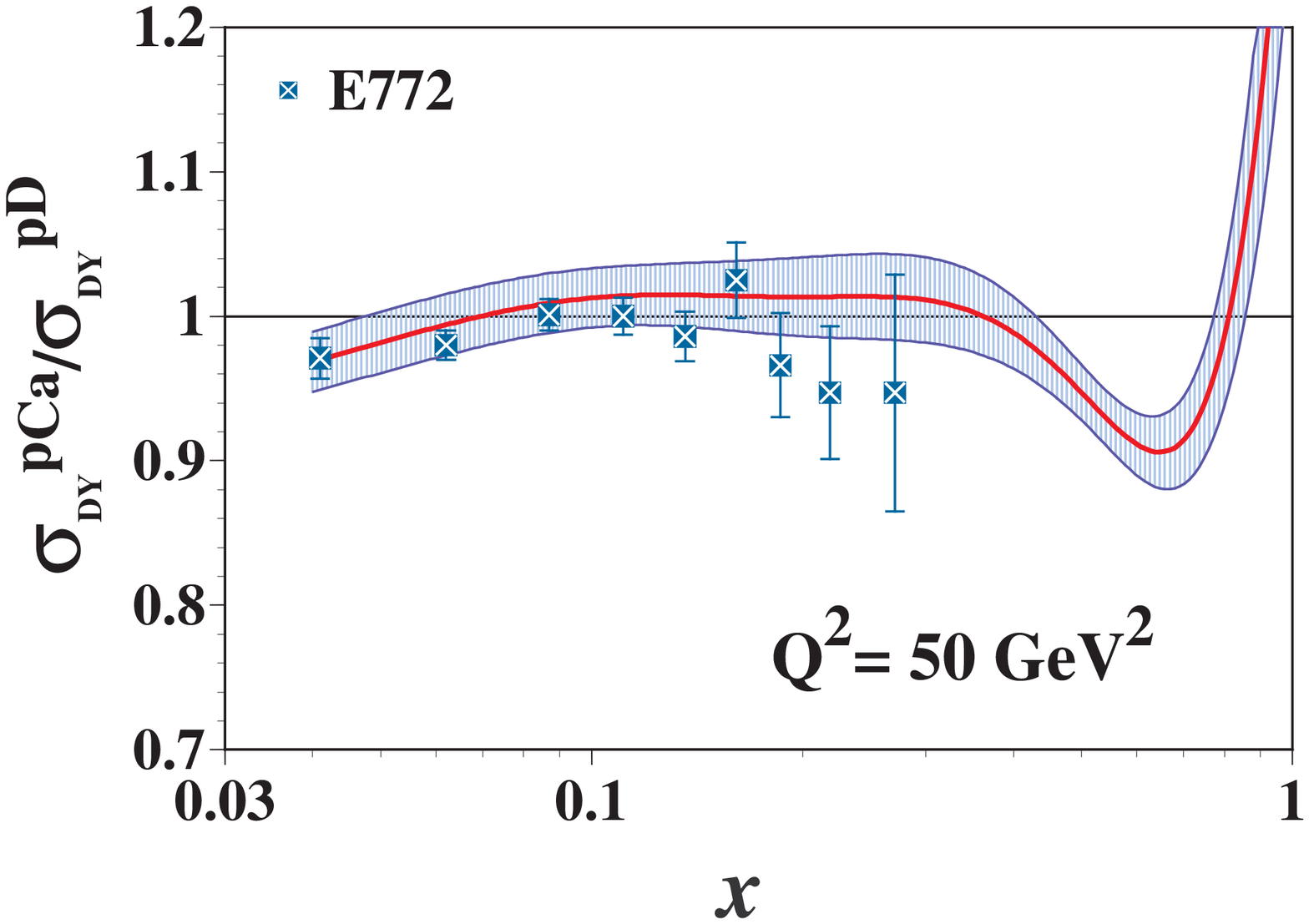} 
        \vspace{-0.3cm}
\caption{(Color online) Parametrization results are compared with 
         the data of $F_2$ ratios $F_2^{Ca}/F_2^{D}$
         and Drell-Yan ratios
         $\sigma_{DY}^{pCa}/\sigma_{DY}^{pD}$.
         The theoretical curves and uncertainties are calculated 
         at $Q^2$=5 GeV$^2$ for the $F_2$ ratios and 
         at $Q^2$=50 GeV$^2$ for the Drell-Yan ratios.}
\label{fig:f2-dy}
\vspace{-0.3cm}
\end{figure}

In the actual fit, the parameters for the Fermi-motion part are fixed at
$\beta_v$=$\beta_{\bar q}$=$\beta_g$=0.1 because of the lack of large-$x$ data.
The parameter $\alpha$ is also fixed at $\alpha=1/3$ \cite{saga01}
for the $A$ dependence. 
The parameters obtained by the $\chi^2$ analysis are shown in
Table \ref{table:parameters}. 
Three parameters are fixed by the charge, baryon-number, and momentum
conservations, and they are chosen $a_{u_v}$, $a_{d_v}$, and $a_{g}$
in the analysis. Because these constants depend on nuclear species,
they are listed separately in Appendix \ref{appen-a}.
Another parameter $c_g$ is also fixed since the gluon parameters
cannot be determined easily by the present data.

The $\chi^2$ analysis results are shown in comparison with the data.
First, $\chi^2$ values are listed for each nuclear data set in
Table \ref{tab:chi2}. The total $\chi^2$ divided by the degree of freedom
is 1.58. Comparison with the actual data is shown in Figs.
\ref{fig:RD}, \ref{fig:RA}, and \ref{fig:DY} for the $F_2^A/F_2^D$,
$F_2^A/F_2^{C,Li}$, and Drell-Yan ($\sigma_{DY}^{pA}/\sigma_{DY}^{pA'}$)
data, respectively. These ratios are denoted $R^{exp}$ for the experimental
data and $R^{theo}$ for the parametrization calculations. The deviation
ratios $(R^{exp}-R^{theo})/R^{theo}$ are shown in these figures.
The NPDFs are evolved to the experimental $Q^2$ points,
then the ratios $(R^{exp}-R^{theo})/R^{theo}$ are calculated. 

As examples, actual data are compared with the parametrization
results in Fig. \ref{fig:f2-dy} for the ratios $F_2^{Ca}/F_2^{D}$ 
and $\sigma_{DY}^{pCa}/\sigma_{DY}^{pD}$. The shaded areas indicate
the ranges of NPDF uncertainties, which are calculated at $Q^2$=5 GeV$^2$
for the $F_2$ ratios and at $Q^2$=50 GeV$^2$ for the Drell-Yan ratios.
The experimental data are well reproduced by the parametrization,
and the the data errors agree roughly with the uncertainty bands.
We should note that the parametrization curves and the uncertainties
are calculated at at $Q^2$=5 and 50 GeV$^2$, whereas the data are taken
at various $Q^2$ points. In Fig. \ref{fig:f2-dy}, the smallest-$x$ data 
at $x$=0.0062 for $F_2^{Ca}/F_2^{D}$ seems to deviate from
the parametrization curve. However, the deviation
comes simply from a $Q^2$ difference. In fact, if the theoretical ratio
is estimated at the experimental $Q^2$ point, the data point
agrees with the parametrization as shown in Fig. \ref{fig:RD}.

In general, the figures indicate a good fit to the data, which
suggests that the $\chi^2$ analysis should be successful.
However, there are some deviations as indicated in the table and figures.
The $\chi^2$ contributions are large from small nuclei.
For example, the $Li/D$ ratios have the $\chi^2$ value 88.7
for only 17 data points. In fact, the $Li/D$ ratios in the region
$x \sim 0.01$ deviate from the theoretical curve in Fig. \ref{fig:RD}.
The $Li/D$ ratios are measured with small errors so that
they produce large $\chi^2$ values.
However, if we wish to reproduce the $Li/D$ ratios, the $^4 He/D$ and
$Be/D$ ratios cannot be well explained. This is why the {\tt MINUIT}
subroutine produced the optimum point although theoretical calculations
deviate from the experimental $Li/D$ ratios. We also notice that
the $Sn/C$, $C/Li$, and $Ca/Li$ ratios are not well reproduced
in the region $x < 0.04$. On the other hand, the figures indicate
that medium- and large-size nuclei are well explained
by the parametrization model.

The Drell-Yan data are taken mainly in the range $0.02<x<0.2$ as shown
in Fig. \ref{fig:DY}. The Drell-Yan cross section ratio
$\sigma_{DY}^{pA}/\sigma_{DY}^{pA'}$ is almost identical to
the antiquark ratio $\bar q^A(x_2)/\bar q^{A'}(x_2)$
in the $x$ region, $x<0.1$. Therefore, the Drell-Yan data are
especially valuable for determining the antiquark modification
in the $x$ region, $x\sim 0.1$. In the smaller $x$ region, the antiquark
shadowing is fixed by the $F_2$ data in any case. 
Except for the $W/Be$ Drell-Yan ratios in the region $x \sim 0.02$,
the data are well explained by the parametrization.
From the constraints of these Drell-Yan cross sections, $F_2$ shadowing,
and momentum conservation, the antiquark distributions are relatively well
determined in the region $0.006 < x < 0.1$.
However, the behavior of the medium- and large-$x$ regions is not obvious. 

\subsection{Comparison with $Q^2$-dependent data}
\label{comp-q2}

\begin{figure}[b!]
\vspace{-0.3cm}
\includegraphics[width=0.42\textwidth]{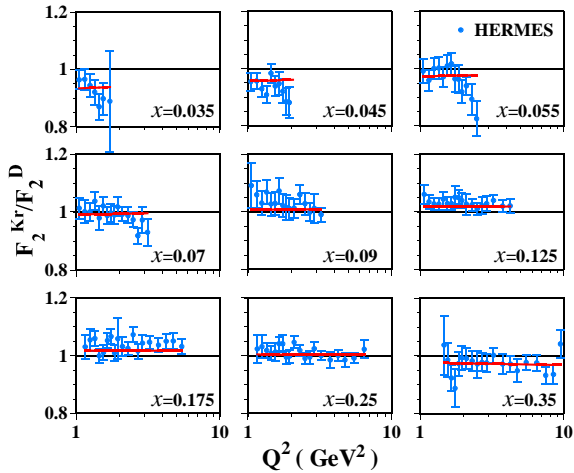}
\vspace{-0.2cm}
\caption{(Color online) $Q^2$ dependence of $F_2^{Kr}/F_2^{D}$.
         The curves indicate parametrization results.}
\label{fig:krddy}
\end{figure}
\vspace{-0.0cm}
\begin{figure}[h]
\includegraphics[width=0.42\textwidth]{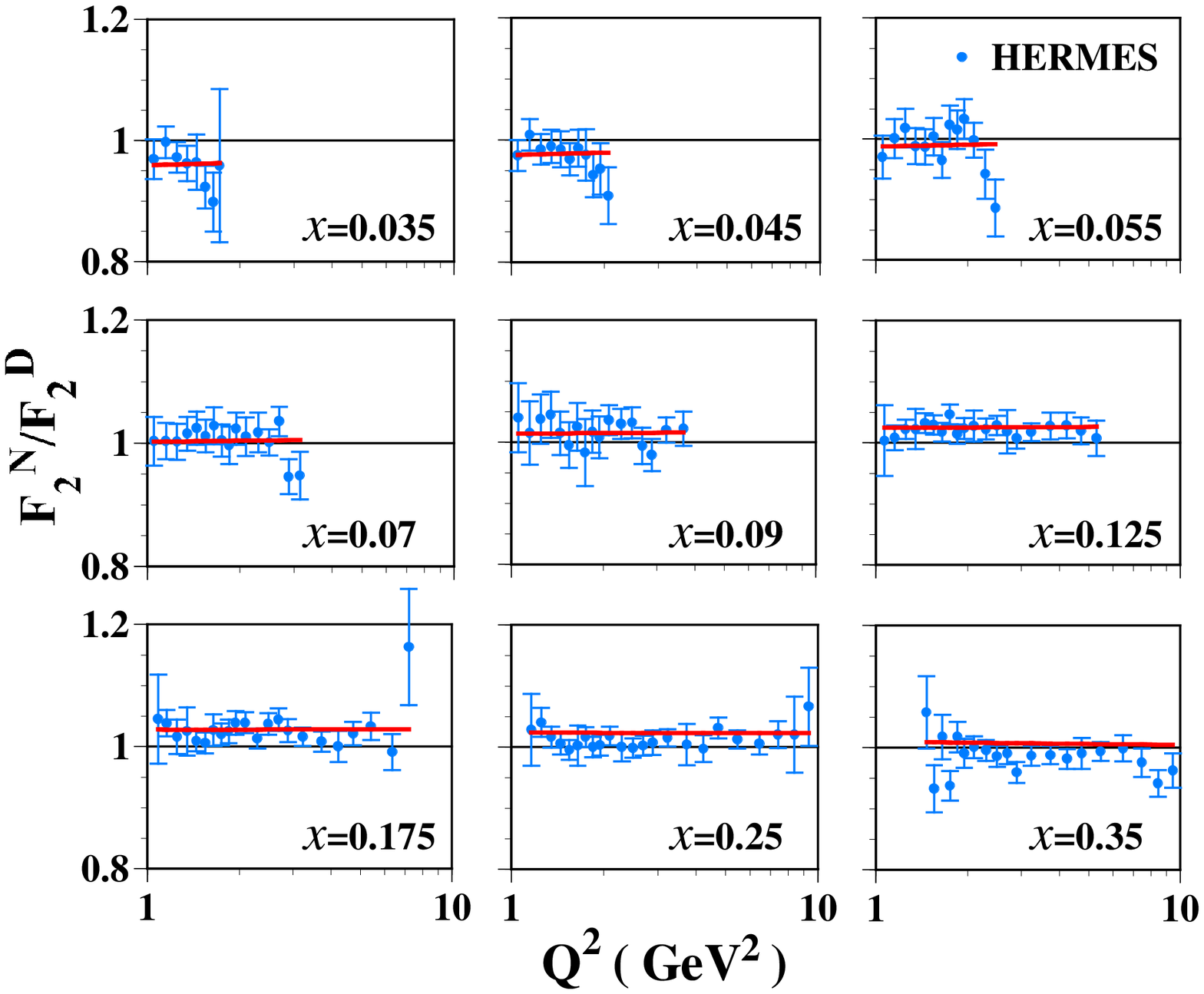}
\vspace{-0.2cm}
\caption{(Color online) $Q^2$ dependence of $F_2^{N}/F_2^{D}$.}
\label{fig:nddy}
\end{figure}
\vspace{-0.0cm}
\begin{figure}[h]
\vspace{-0.0cm}
\includegraphics[width=0.42\textwidth]{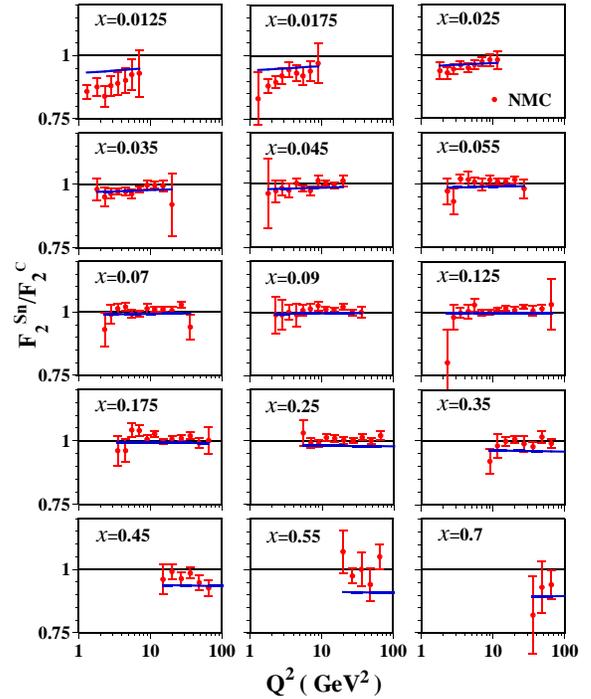}
\vspace{-0.2cm}
\caption{(Color online) $Q^2$ dependence of $F_2^{Sn}/F_2^{C}$.}
\label{fig:snddy}
\vspace{-0.2cm}
\end{figure}

The analysis results are compared with $Q^2$ dependent data in Figs.
\ref{fig:krddy}, \ref{fig:nddy}, and \ref{fig:snddy}
for the ratios,
$F_2^{Kr}/F_2^{D}$, $F_2^{N}/F_2^{D}$, and  $F_2^{Sn}/F_2^{C}$,
respectively. The fit results are shown by the curves in these
figures. The data are well reproduced by the fit except for the
$Sn/C$ ratios at small and medium $x$. The tin shadowing is underestimated 
in comparison with the carbon shadowing as indicated in the previous
subsection. However, we notice that the experimental data are not
``consistent" in the sense that the $F_2^{Kr}/F_2^{D}$ and $F_2^{N}/F_2^{D}$
ratios tend to decrease at $x$=0.035 and 0.045 with increasing $Q^2$, whereas
the $F_2^{Sn}/F_2^{C}$ ratio increases. Obviously, more detailed
experimental investigations should be done for clarifying
the $Q^2$ dependence. It is especially important for fixing the gluon
distributions in nuclei. The $Q^2$ dependence is related partially
to the nuclear gluon distributions through the $Q^2$ evolution
equations. If the experimental $Q^2$ dependence becomes
clear, we should be able to pin down the nuclear gluon modification.

\subsection{Optimum parton distribution functions}
\label{dist}

\begin{figure}[b]
        \includegraphics*[width=40mm]{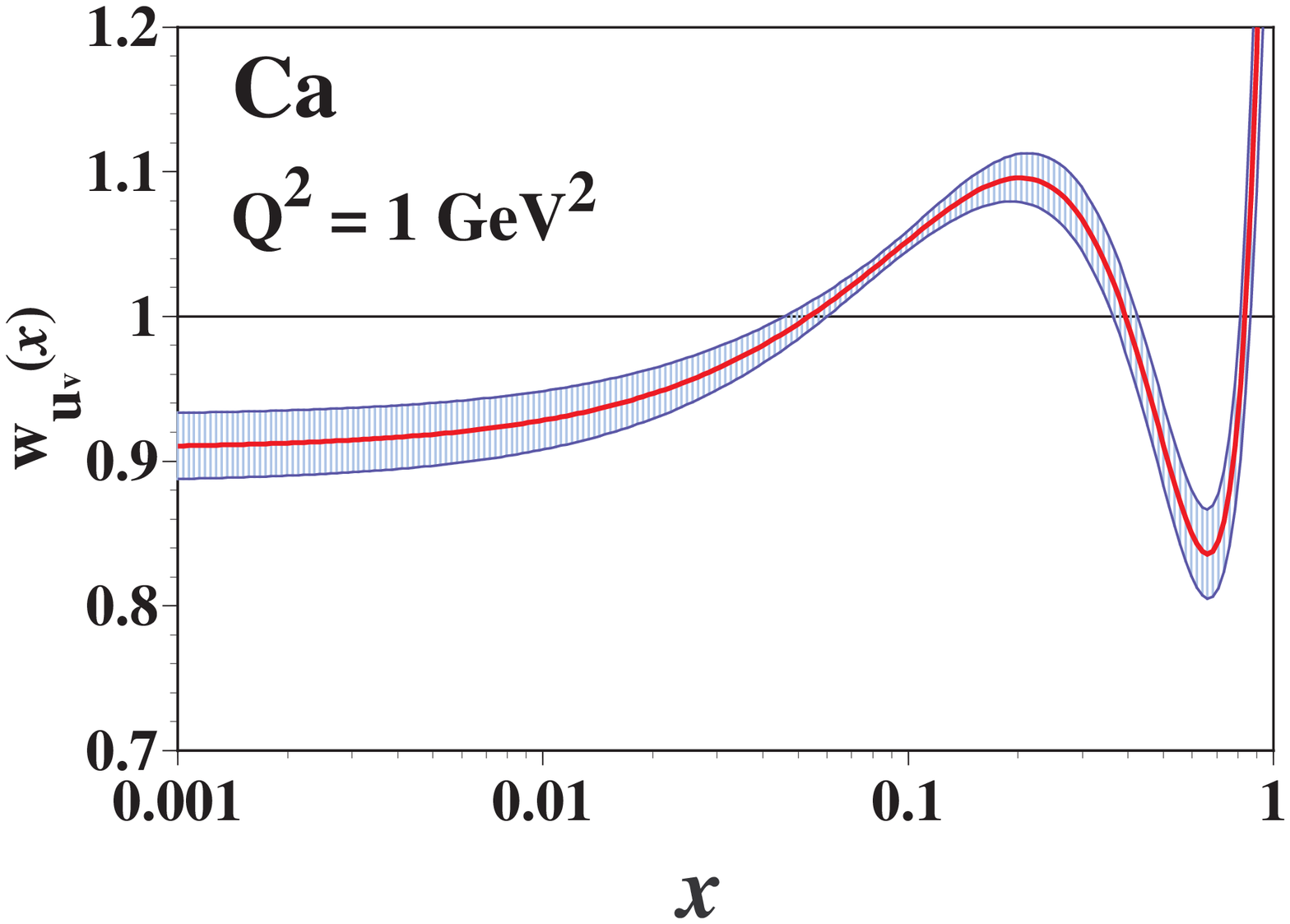} \hspace{0.5mm}
        \includegraphics*[width=40mm]{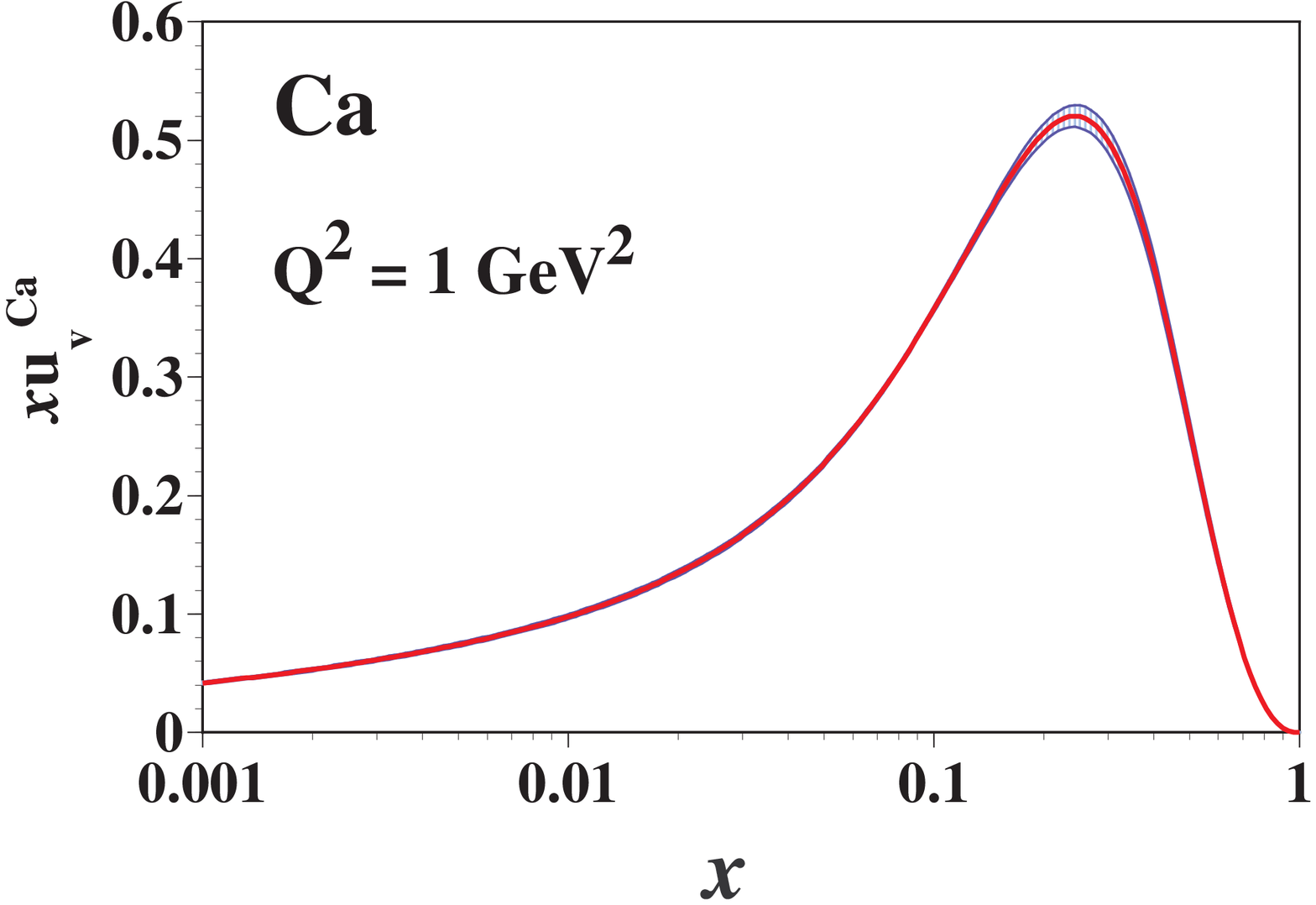} \\
\vspace{0.5cm}
        \includegraphics*[width=40mm]{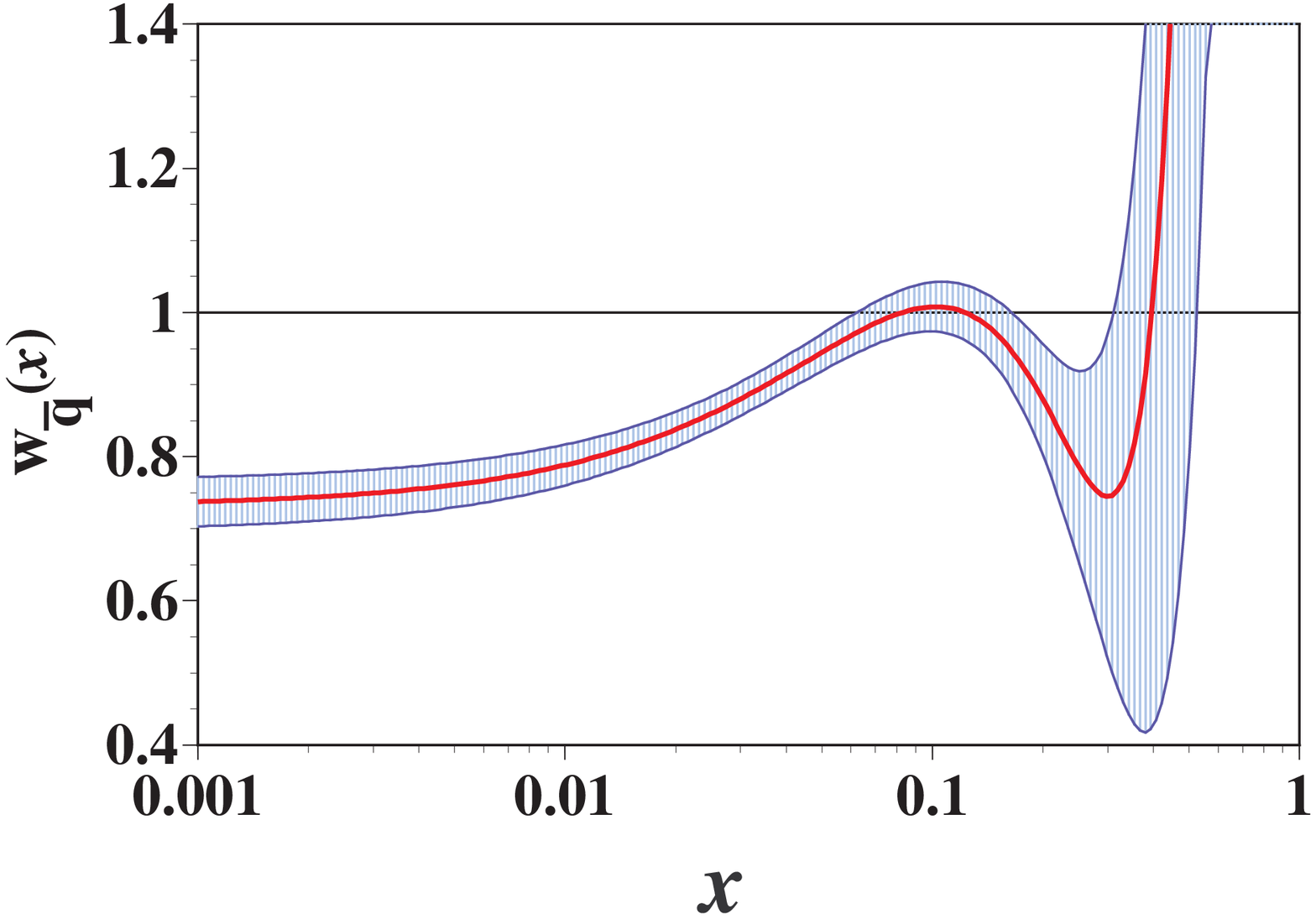} \hspace{0.5mm}
        \includegraphics*[width=40mm]{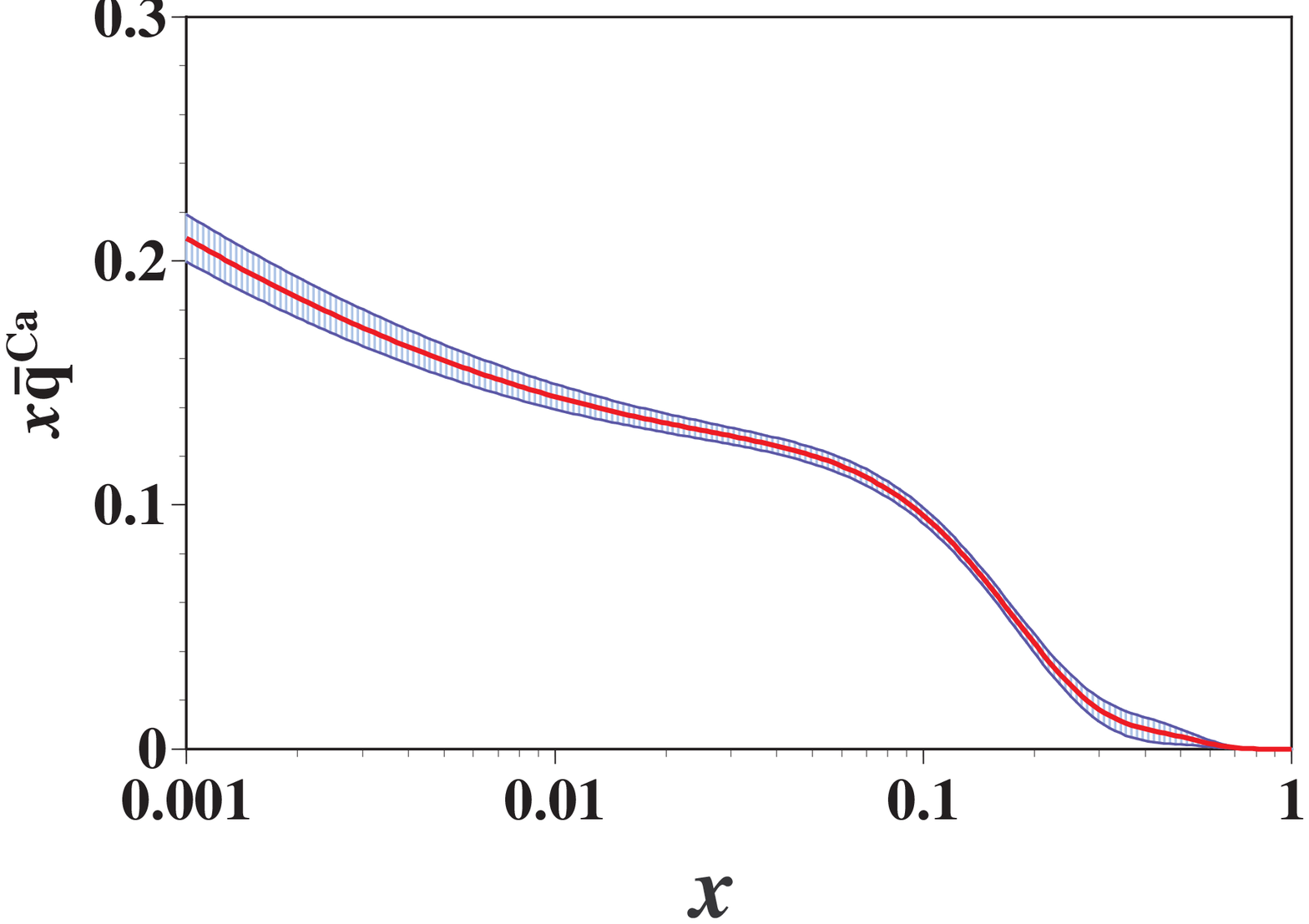} \\
\vspace{0.5cm}
        \includegraphics*[width=40mm]{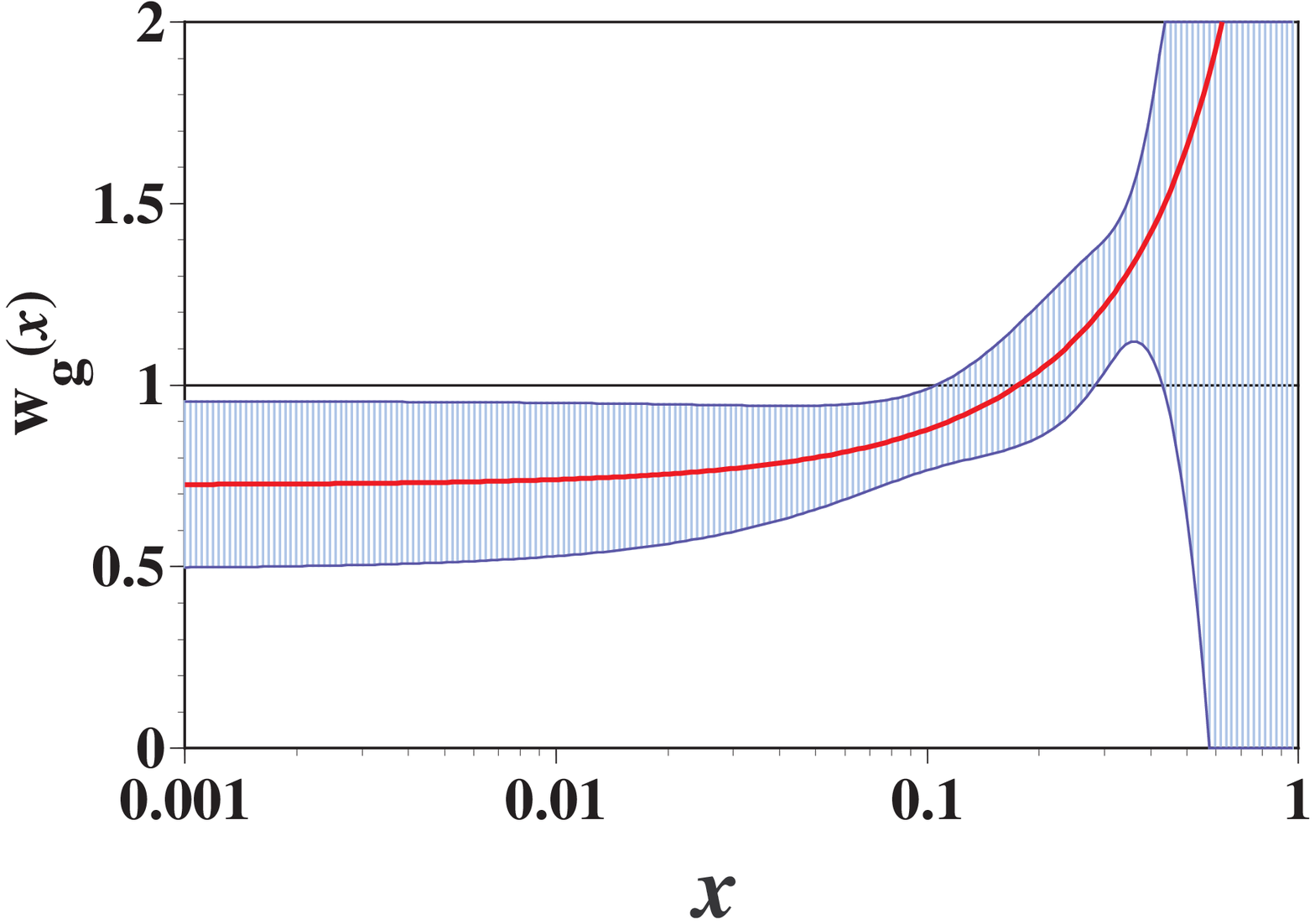} \hspace{0.5mm}
        \includegraphics*[width=40mm]{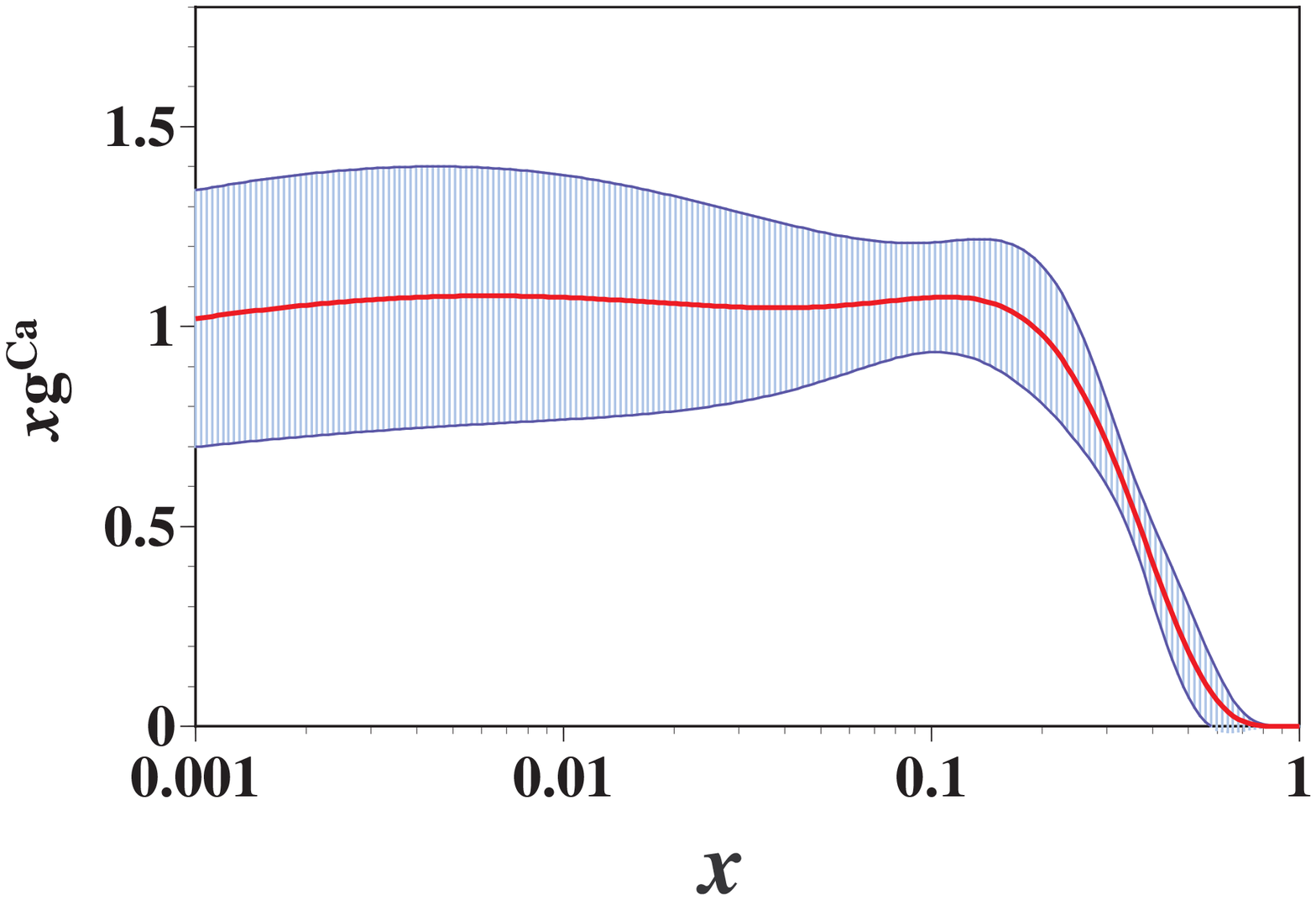} 
       \vspace{-0.0cm}
\caption{(Color online) 
The weight functions and the nuclear parton distribution functions
are shown for the calcium nucleus at $Q_0^2$. The uncertainties
are shown by the bands.
}
\label{fig:npdfs}
\end{figure}

We show the nuclear parton distribution functions obtained by
the $\chi^2$ analysis. As a typical medium-sized nucleus, the calcium
is selected for showing the distributions. Because it is an isoscalar
nucleus, the $u_v^A$ and $d_v^A$ are identical. Therefore, 
$u_v^{Ca}$ (=$d_v^{Ca}$),  $\bar q ^{\, Ca}$, and $g^{Ca}$ and
their weight functions are shown in Fig. \ref{fig:npdfs}
at $Q_0^2$.

The valence-quark modification $w_{u_v}$ is precisely determined by the data
in the medium- and large-$x$ regions. However, the uncertainty band
becomes larger in the region $x<0.03$ although it is constrained somewhat
by the charge and baryon-number conservations. Obviously, we should
wait for NuMI (Neutrinos at the Main Injector) \cite{numi} and
neutrino-factory \cite{nufact03} projects for clarifying
the valence-quark shadowing by the structure function $F_3$.
Although the uncertainties of the nuclear modification $w_{u_v}^{Ca}$
are relatively large at $x<0.03$, it is not so obvious in the valence-quark
distribution ($x u_v^{Ca}$), as shown on the right-hand side of
Fig. \ref{fig:npdfs}, because the distribution is small in
the small-$x$ region.
 
We should mention the possibility that the uncertainties could be
underestimated because we fixed some parameters such as $\alpha$ and
$\beta$ in the analysis. In addition, there should be uncertainties 
from the assumed functional form. These additional factors will be
investigated in future. In this respect, it is certainly worth while
investigating the $F_3$ shadowing at future neutrino facilities
\cite{numi,nufact03} in spite of the analysis result for the
valence-quark shadowing in Fig. \ref{fig:npdfs}.

The uncertainties of the antiquark modification $w_{\bar q}^{Ca}$ are
small in the region $x<0.1$ because it is fixed by the $F_2$ and 
Drell-Yan data. However, it has large uncertainties in
the $x$ region, $x>0.2$. The antiquark distribution $x \bar q^{\, Ca}$
itself is small at $x>0.2$, so that it becomes difficult to take accurate
data for the nuclear modification. In order to determine the distribution
in this region, we need another Drell-Yan experiment which is intended
especially for large-$x$ physics \cite{dy-large-x}. 

\begin{figure}[b]
        \includegraphics*[width=40mm]{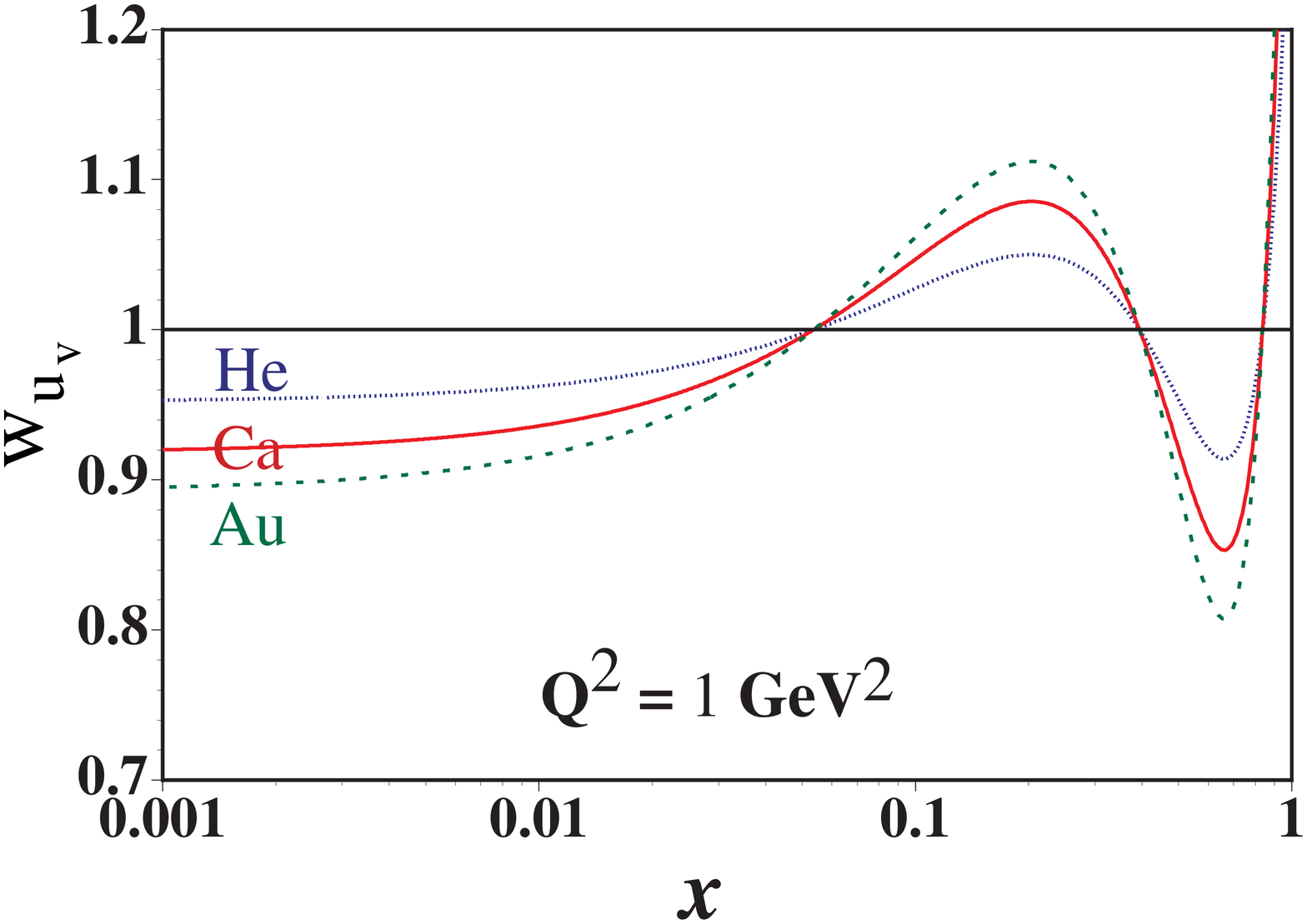} \hspace{0.5mm}
        \includegraphics*[width=40mm]{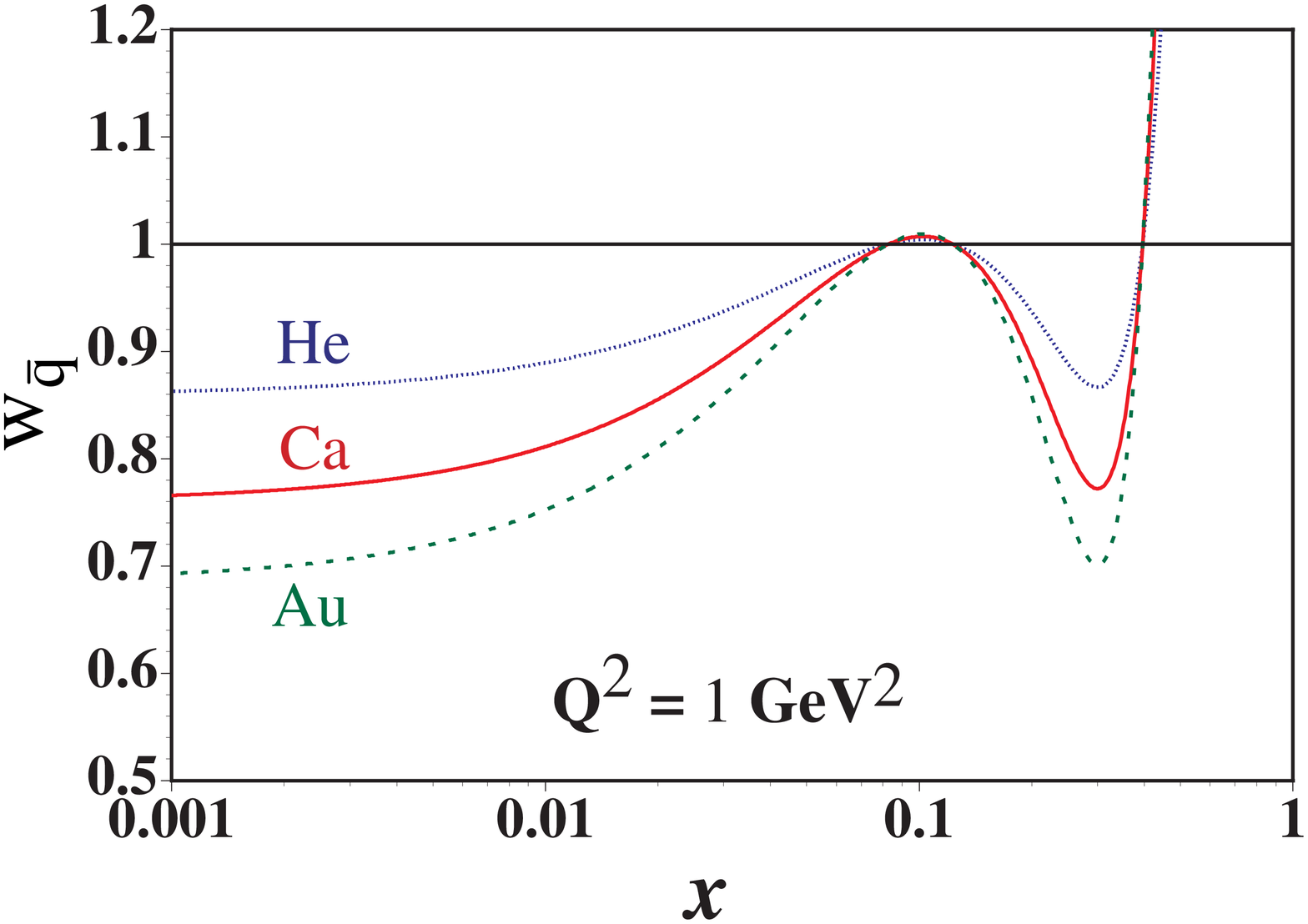} \\
\vspace{0.2cm}
        \includegraphics*[width=40mm]{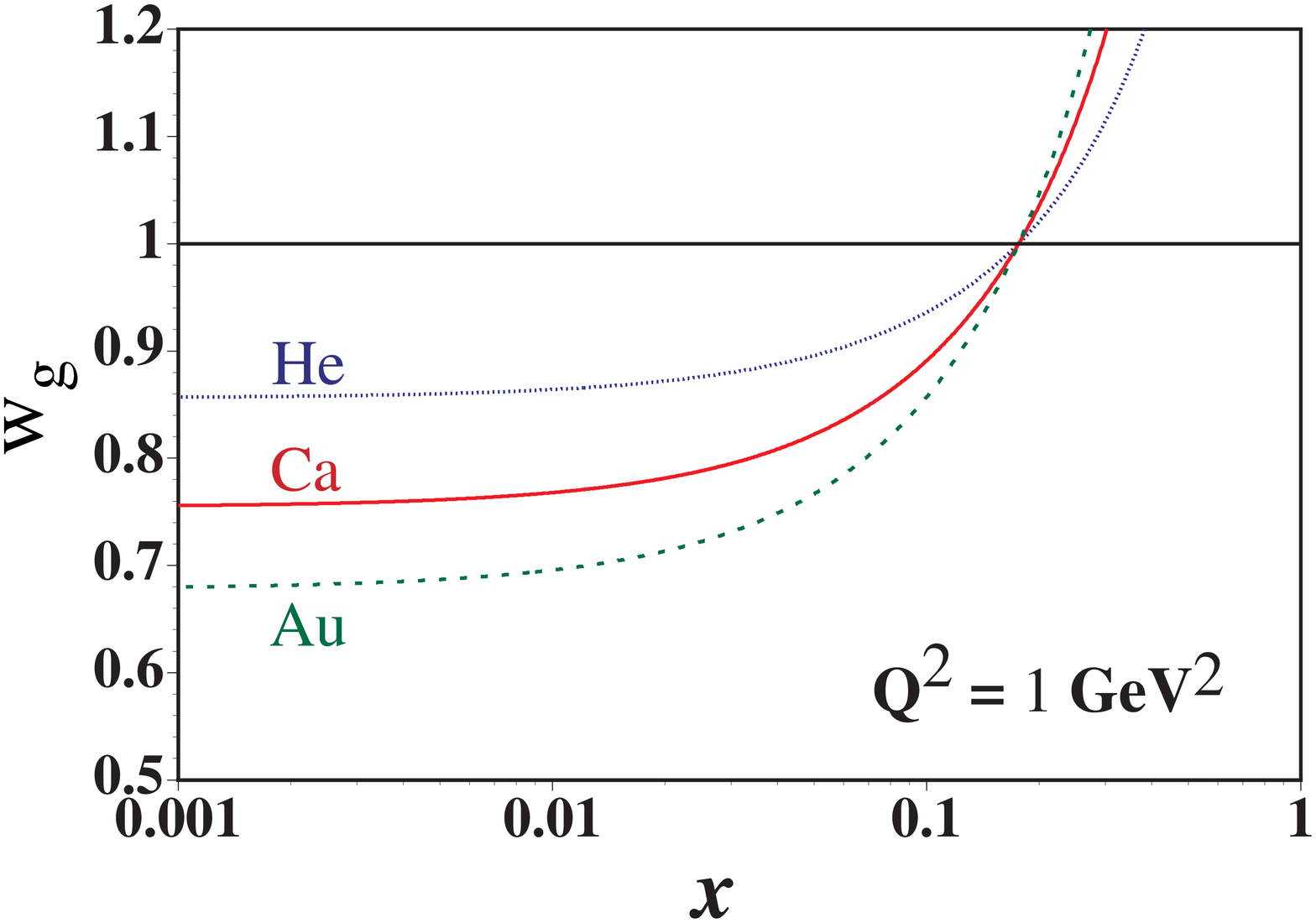}
        \vspace{-0.2cm}
\caption{(Color online) 
The weight functions are shown for the nuclei, $^4$He, Ca, and Au,
at $Q_0^2$.
}
\label{fig:npdfs-a}
\end{figure}

The gluon distribution is especially difficult to be determined by
the present data. It is clearly shown in Fig. \ref{fig:npdfs} that
the modification $w_{g}^{Ca}$ and the distribution $xg^{Ca}$ have
large uncertainties. As explained in the previous subsection,
the nuclear $Q^2$ dependence is not clear from the data. This fact
makes it difficult to fix the nuclear gluon distributions.
However, we notice that the gluon distribution seems to be shadowed
although the uncertainties are large at $x<0.1$.

We notice that the functional form of the gluon weight function $w_g$
is different from those of the valence-quark and antiquark functions,
$w_v$ and $w_{\bar q}$. A similar functional form was also tested
in the analysis. We provided a weight function $w_g$, which has the same
functional form with $w_{\bar q}$, as the initial one for the $\chi^2$
analysis without fixing the parameter $c_g$. However, the analysis
ended up with gluon distributions which are similar to the one
in Fig. \ref{fig:npdfs}. It is simply because of the lack of data which
are sensitive to the gluon distributions. It is the reason why we
decided to fix the parameter $c_g$ in the current analysis.
The gluon distributions play an important role in many aspects of
high-energy heavy-ion collisions, so that they should be determined
by future experimental data. For example, the eRHIC project \cite{erhic}
could be a promising one for determining the nuclear PDFs at small $x$.
In order to illustrate the nuclear dependence of the PDFs, we show
the weight functions for the nuclei, $^4$He, Ca, and Au,
in Fig. \ref{fig:npdfs-a} 

For general users, a computer code is available on the web site
\cite{nucl-lib} for calculating the parton distribution
functions for nuclei at given $x$ and $Q^2$. 
The details are explained in Appendix B.

\section{Summary}\label{summary}

The nuclear parton distribution functions and their uncertainties are
determined by analyzing the experimental data of $F_2$ and Drell-Yan data.
The uncertainties are estimated by the Hessian method.
The valence-quark distributions are well determined except for
the region $x<0.03$. The antiquark distributions have small uncertainties
at $x<0.1$; however, they cannot be fixed in the region, $x>0.2$.
The gluon distributions have large uncertainties in the whole-$x$ region.
Obviously, we need much accurate scaling violation data or other ones
for fixing the gluon distributions in nuclei.

\begin{acknowledgments}
The authors thank A. Br\"ull for providing the HERMES data \cite{hermes03}
and thank J.-C. Peng and M. A. Vasiliev for the Fermilab-E772/E866-NuSea
data \cite{e772-90,e866-99}. They also thank M.-A. Nakamura for discussions.
S.K. was supported by the Grant-in-Aid for Scientific Research from
the Japanese Ministry of Education, Culture, Sports, Science, and Technology.
\end{acknowledgments}

\appendix
\section{Nuclear dependent parameters}
\label{appen-a}

In Table \ref{table:parameters}, the constants $a_{u_v}$, $a_{d_v}$,
and $a_{\bar q}$ are not listed. These constants are fixed by the three
conservation equations, so that they depend on the mass number $A$
and the atomic number $Z$. For practical usage, we express these
constants by eight integral values $I_{1-8}$ as explained
in Ref. \cite{saga01}:
\begin{align}
a_{u_v}(A,Z) & = - \frac{Z I_1 + (A-Z) I_2}{Z I_3 + (A-Z) I_4} ,
\nonumber \\
a_{d_v}(A,Z) & = - \frac{Z I_2 + (A-Z) I_1}{Z I_4 + (A-Z) I_3} ,
\nonumber \\
a_{g}(A,Z)   & = - \frac{1}{I_8} \,  \bigg[ \,
            a_{u_v}(A,Z) \left\{ \frac{Z}{A} I_5
                                +\left(1-\frac{Z}{A} \right) I_6 \right\}
\nonumber \\
         & \! \! \! \! \! \! \! 
           +a_{d_v}(A,Z) \left\{ \frac{Z}{A} I_6
                                +\left(1-\frac{Z}{A} \right) I_5 \right\}
           +I_7 \, \bigg]   .
\label{eqn:aconst}
\end{align}

Values of the integrals are listed in Table \ref{tab:aconst}
from the current analysis. Using these values together with
Eq. (\ref{eqn:aconst}), one could calculate the constants 
$a_{u_v}$, $a_{d_v}$, and $a_{\bar q}$ for any nucleus. Then, it is
possible to express the nuclear parton distribution functions
analytically at $Q_0^2$ for a given nucleus together
with the MRST01 distributions \cite{mrst01} in the nucleon.

\begin{table}[h]
\caption{Values of the eight integrals.}
\label{tab:aconst}
\begin{ruledtabular}
\begin{tabular*}{\hsize}
{c@{\extracolsep{0ptplus1fil}}c|@{\extracolsep{0ptplus1fil}}c
@{\extracolsep{0ptplus1fil}}c}
Integral & Value     \ \ \ \ \ \ \ \ \    & Integral & Value       \\
\colrule
$I_1$    &  0.2611  \ \ \ \ \ \ \ \ \    & $I_5$    &  0.3445     \\
$I_2$    &  0.1313  \ \ \ \ \ \ \ \ \    & $I_6$    &  0.1345     \\
$I_3$    &  2.018   \ \ \ \ \ \ \ \ \    & $I_7$    &  0.2162     \\
$I_4$    &  1.016   \ \ \ \ \ \ \ \ \    & $I_8$    &  0.3969      
\end{tabular*}
\end{ruledtabular}
\end{table}

\section{Practical code for calculating nuclear PDFs}
\label{appen-b}

One could calculate nuclear PDFs by using the information provided
in Appendix A and in Table \ref{table:parameters}. However,
the distributions should be evolved if one wish to obtain them
at different $Q^2$. For those who are not familiar
with such $Q^2$ evolution, we prepared a practical code for calculating
the nuclear PDFs at given $x$ and $Q^2$. The code could be obtained from
the web site in Ref. \cite{nucl-lib}.

Instructions for using the code are provided in the package. 
Only restrictions are the kinematical ranges, $10^{-9} \le x \le 1$,
and 1 GeV$^2 \le Q^2 \le 10^8$ GeV$^2$. The largest nucleus in the analysis
is the lead, so that it is suitable to use the code within the range
$A \le 208$. However, variations of the NPDFs are rather small from
$A=208$ to the nuclear matter, one could possibly use the code also for
large nuclei with $A>208$.
In the NPDF library, we provide the distributions at very small $x$
as small as $10^{-9}$ for those who use them in integrating 
the distributions over the wide range of $x$. However, one should
be careful that the distributions are not reliable in the region
$x<0.006$, where no experimental data exists. Furthermore,
there is a possibility that higher-twist effects could alter
the results in the small-$x$ region.

The analysis was made in the region, $Q^2 \ge$1 GeV$^2$, where
the perturbative QCD is considered to be applicable.
The obtained NPDFs can be used for high-energy nuclear reactions
with $Q^2 \ge$1 GeV$^2$. However, there are data which are slightly
below this region. For example, many long-baseline neutrino data
are taken in the smaller $Q^2$ region. A useful parametrization
was proposed to describe the cross section from the deep inelastic
region to the resonance one \cite{by02}. We could possibly make
a similar analysis in future for describing lepton-nucleus cross
sections also in the resonance region.



\end{document}